\documentclass[prl,aps,
twocolumn,
nofootinbib,
floatfix,
superscriptaddress]{revtex4}

\usepackage{amsmath,amssymb}
\usepackage{mathrsfs}
\usepackage{psfrag}
\usepackage{hyperref}
\usepackage{wrapfig}
\usepackage{color}

\usepackage{slashed}

\usepackage{simplewick}
\usepackage{cancel}

\usepackage{todonotes}
\usepackage{tikz, pict2e}

\usepackage{lmodern}
\usepackage{tkz-euclide}
\usepackage[utf8]{inputenc}

\usepackage{float}

\usetikzlibrary{calc,backgrounds, intersections}
\usetikzlibrary{arrows,shapes,positioning,shadows,trees}
\usetikzlibrary{mindmap}
\usetikzlibrary{decorations.pathreplacing}
\usetikzlibrary{decorations.pathmorphing}
\usetikzlibrary{decorations.markings}
\usetikzlibrary{matrix}

\tikzstyle arrowstyle=[scale=1]
\tikzstyle directed=[postaction={decorate,decoration={markings,
    mark=at position .5 with {\arrow[arrowstyle]{stealth}}}}]
\tikzstyle reverse directed=[postaction={decorate,decoration={markings,
    mark=at position .5 with {\arrowreversed[arrowstyle]{stealth};}}}]

\usepackage{mathtools,bbm,array,amsfonts,graphicx,wrapfig,arydshln,lscape,float,multirow,longtable,rotating,makecell}
\usepackage{url}

\definecolor{redb}{rgb}{0.700, 0.000, 0.300}
\DeclareMathAlphabet\mathbfcal{OMS}{cmsy}{b}{n}

\def\r{\rho}

\def\aa1{\phi}
\def\cc1{\psi}

\catcode`\@=12

\begin{document}


\title{\small On the Emergence of Lorentz Invariance and Unitarity from the Scattering Facet of Cosmological Polytopes}

\author{\small Nima Arkani-Hamed}\affiliation{\small School of Natural Sciences, Institute for Advanced Study, Princeton, NJ}
\author{\small Paolo Benincasa}\affiliation{\small Niels Bohr International Academy and Discovery Center, Copenhagen, Denmark}

\begin{abstract}
The concepts of Lorentz invariance of local (flat space) physics, and unitarity of time evolution and the S-matrix, are famously rigid and robust, admitting no obvious consistent theoretical deformations, and confirmed to incredible accuracy by experiments. But neither of these notions seem to appear directly in describing the spatial correlation functions at future infinity characterizing the ``boundary" observables in cosmology. How then can we see them emerge as {\it exact} concepts from a possible ab-initio theory for the late-time wavefunction of the universe? In this letter we examine this question in a simple but concrete setting, for the perturbative wavefunction in a class of scalar field models where an ab-initio description of the wavefunction has been given by ``cosmological polytopes". Singularities of the wavefunction are associated with facets of the polytope. One of the singularities -- corresponding to the ``total energy pole" -- is well known to be associated with the flat-space scattering amplitude. We show how the combinatorics and geometry of this {\it scattering facet} of the cosmological polytope straightforwardly leads to the emergence of Lorentz invariance and unitarity for the S-matrix.  Unitarity follows from the way boundaries of the scattering facet factorize into products of lower-dimensional polytopes, while Lorentz invariance follows from a contour integral representation of the canonical form, which exists for any polytope, specialized to cosmological polytopes.
\end{abstract}

\maketitle

\section{Introduction}

Lorentz invariance in flat space, together with quantum-mechanical unitarity of time evolution, are the foundations of fundamental physics. There are countless indications that these concepts are totally rigid and cannot be deformed. It is very unlikely that Lorentz invariance is emergent from some microscopic system with a preferred frame since this would infect the dimensionless couplings (such as different species ``speeds of light''), while all modifications of the known rules of quantum mechanics are either wildly inconsistent or run afoul of the locality demanded by special relativity. Indeed, these are some of the indications that space-time and quantum mechanics are tied to each other in some deep way.

On the other hand, at least naively, cosmology suggests that both of these ideas can somehow {\it not} be fundamental. Of course, Lorentz invariance is broken on cosmological scales, and ultimately the accelerating universe seems to make quantum-mechanical observables of any kind approximate. Even more prosaically, the wavefunction of the universe is a static quantity, depending only on spatial co-ordinates on the future spatial boundary of the universe, and there should be some rules for determining it. Why should such rules know anything about Lorentz invariance (since this object is not Lorentz invariant) or unitarity (since there is no time evolution here)? 

There is a more precise version of this question in perturbation theory. As we will review, the (integrand of the) wavefunction shows a pole in the sum $E_{\mbox{\tiny tot}}$ of the energies of the external states, whose residue is exactly the flat-space scattering amplitude \cite{Raju:2012zr, Raju:2012zs, Arkani-Hamed:2015bza}. Thus, whatever the putative new rules are for determining the wavefunction without referring to either flat-space Lorentz invariance or unitary time-evolution, they must somehow magically produce an object which is {\it exactly} Lorentz-invariant and unitary on such a pole.

In this note we will see exactly how this happens for a wide class of theories of scalar fields with polynomial interactions, where such an ab-initio understanding of the wavefunction has been found in terms of {\it cosmological polytopes}, with a simple, intrinsic definition making no reference to space-time or Hilbert space notions \cite{Arkani-Hamed:2017fdk}. Cosmological polytopes represent a small first step in the direction of finding the analog of objects such as amplituhedra \cite{Arkani-Hamed:2013jha} and associahedra \cite{Arkani-Hamed:2017mur, Frost:2018djd, Salvatori:2018aha}, seen in the context of scattering amplitudes, in cosmology.

As we will see, one of the facets of the polytope is naturally associated with the $E_{\mbox{\tiny tot}}$ pole and should give scattering amplitudes. Indeed, the scattering amplitude plays an important role in the full wavefunction. On {\it any} of its poles, the wavefunction factorizes into a lower-point scattering amplitude times a lower point wavefunction, a fact we can trivially see from the facet structure of the cosmological polytope.

We will study this {\it scattering facet} in detail. At tree-level, as observed already in \cite{Arkani-Hamed:2017fdk}, this facet is just a simplex; remarkably, the canonical form pairs up linear energy poles into the quadratic factors we associate with Lorentz-invariant propagators.

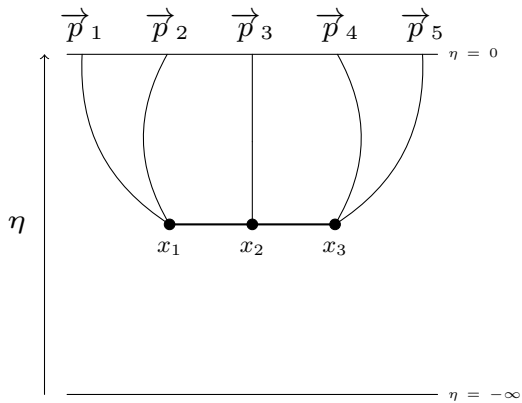
\begin{figure*}
 \centering
 \begin{tikzpicture}[line join = round, line cap = round, ball/.style = {circle, draw, align=center, anchor=north, inner sep=0}, scale=2, transform shape]
  \begin{scope}
   \def\cx{1.75}
   \def\cy{3}
   \def\r{.6}
   \pgfmathsetmacro\Axi{\cx+\r*cos(135)}
   \pgfmathsetmacro\Ayi{\cy+\r*sin(135)}
   \pgfmathsetmacro\Axf{\Axi+cos(135)}
   \pgfmathsetmacro\Ayf{\Ayi+sin(135)}
   \coordinate (pAi) at (\Axi,\Ayi);
   \coordinate (pAf) at (\Axf,\Ayf);
   \pgfmathsetmacro\Bxi{\cx+\r*cos(45)}
   \pgfmathsetmacro\Byi{\cy+\r*sin(45)}
   \pgfmathsetmacro\Bxf{\Bxi+cos(45)}
   \pgfmathsetmacro\Byf{\Byi+sin(45)}
   \coordinate (pBi) at (\Bxi,\Byi);
   \coordinate (pBf) at (\Bxf,\Byf);
   \pgfmathsetmacro\Cxi{\cx+\r*cos(-45)}
   \pgfmathsetmacro\Cyi{\cy+\r*sin(-45)}
   \pgfmathsetmacro\Cxf{\Cxi+cos(-45)}
   \pgfmathsetmacro\Cyf{\Cyi+sin(-45)}
   \coordinate (pCi) at (\Cxi,\Cyi);
   \coordinate (pCf) at (\Cxf,\Cyf);
   \pgfmathsetmacro\Dxi{\cx+\r*cos(-135)}
   \pgfmathsetmacro\Dyi{\cy+\r*sin(-135)}
   \pgfmathsetmacro\Dxf{\Dxi+cos(-135)}
   \pgfmathsetmacro\Dyf{\Dyi+sin(-135)}
   \coordinate (pDi) at (\Dxi,\Dyi);
   \coordinate (pDf) at (\Dxf,\Dyf);
   \pgfmathsetmacro\Exi{\cx+\r*cos(90)}
   \pgfmathsetmacro\Eyi{\cy+\r*sin(90)}
   \pgfmathsetmacro\Exf{\Exi+cos(90)}
   \pgfmathsetmacro\Eyf{\Eyi+sin(90)}
   \coordinate (pEi) at (\Exi,\Eyi);
   \coordinate (pEf) at (\Exf,\Eyf);
   \coordinate (ti) at ($(pDf)-(.25,0)$);
   \coordinate (tf) at ($(pAf)-(.25,0)$);
   \draw[->] (ti) -- (tf);
   \coordinate[label=left:{\tiny $\displaystyle\eta$}] (t) at ($(ti)!0.5!(tf)$);
   \coordinate (t0) at ($(pBf)+(.1,0)$);
   \coordinate (tinf) at ($(pCf)+(.1,0)$);
   \node[scale=.5, right=.0125 of t0] (t0l) {\tiny $\displaystyle\eta\,=\,0$};
   \node[scale=.5, right=.0125 of tinf] (tinfl) {\tiny $\displaystyle\eta\,=\,-\infty$};
   \draw[-] ($(pAf)-(.1,0)$) -- (t0);
   \draw[-] ($(pDf)-(.1,0)$) -- ($(pCf)+(.1,0)$);
   \coordinate (d2) at ($(pAf)!0.25!(pBf)$);
   \coordinate (d3) at ($(pAf)!0.5!(pBf)$);
   \coordinate (d4) at ($(pAf)!0.75!(pBf)$);
   \node[above=.01cm of pAf, scale=.625] (d1l) {$\displaystyle\overrightarrow{p}_1$};
   \node[above=.01cm of d2, scale=.625] (d2l) {$\displaystyle\overrightarrow{p}_2$};
   \node[above=.01cm of d3, scale=.625] (d3l) {$\displaystyle\overrightarrow{p}_3$};
   \node[above=.01cm of d4, scale=.625] (d4l) {$\displaystyle\overrightarrow{p}_4$};
   \node[above=.01cm of pBf, scale=.625] (d5l) {$\displaystyle\overrightarrow{p}_5$};
   \def\rb{.55}
   \pgfmathsetmacro\sax{\cx+\rb*cos(180)}
   \pgfmathsetmacro\say{\cy+\rb*sin(180)}
   \coordinate[label=below:{\scalebox{0.5}{$x_1$}}] (s1) at (\sax,\say);
   \pgfmathsetmacro\sbx{\cx+\rb*cos(135)}
   \pgfmathsetmacro\sby{\cy+\rb*sin(135)}
   \coordinate (s2) at (\sbx,\sby);
   \pgfmathsetmacro\scx{\cx+\rb*cos(90)}
   \pgfmathsetmacro\scy{\cy+\rb*sin(90)}
   \coordinate (s3) at (\scx,\scy);
   \pgfmathsetmacro\sdx{\cx+\rb*cos(45)}
   \pgfmathsetmacro\sdy{\cy+\rb*sin(45)}
   \coordinate (s4) at (\sdx,\sdy);
   \pgfmathsetmacro\sex{\cx+\rb*cos(0)}
   \pgfmathsetmacro\sey{\cy+\rb*sin(0)}
   \coordinate[label=below:{\scalebox{0.5}{$x_3$}}] (s5) at (\sex,\sey);
   \coordinate[label=below:{\scalebox{0.5}{$x_2$}}] (sc) at (\cx,\cy);
   \draw (s1) edge [bend left] (pAf);
   \draw (s1) edge [bend left] (d2);
   \draw (s3) -- (d3);
   \draw (s5) edge [bend right] (d4);
   \draw (s5) edge [bend right] (pBf);
   \draw [fill] (s1) circle (1pt);
   \draw [fill] (sc) circle (1pt);
   \draw (s3) -- (sc);
   \draw [fill] (s5) circle (1pt);
   \draw[-,thick] (s1) -- (sc) -- (s5);
  \end{scope}
  \begin{scope}[shift={(4.5,0)}, transform shape]
   \def\cx{1.75}
   \def\cy{3}
   \def\r{.6}
   \pgfmathsetmacro\Axi{\cx+\r*cos(135)}
   \pgfmathsetmacro\Ayi{\cy+\r*sin(135)}
   \pgfmathsetmacro\Axf{\Axi+cos(135)}
   \pgfmathsetmacro\Ayf{\Ayi+sin(135)}
   \coordinate (pAi) at (\Axi,\Ayi);
   \coordinate (pAf) at (\Axf,\Ayf);
   \pgfmathsetmacro\Bxi{\cx+\r*cos(45)}
   \pgfmathsetmacro\Byi{\cy+\r*sin(45)}
   \pgfmathsetmacro\Bxf{\Bxi+cos(45)}
   \pgfmathsetmacro\Byf{\Byi+sin(45)}
   \coordinate (pBi) at (\Bxi,\Byi);
   \coordinate (pBf) at (\Bxf,\Byf);
   \pgfmathsetmacro\Cxi{\cx+\r*cos(-45)}
   \pgfmathsetmacro\Cyi{\cy+\r*sin(-45)}
   \pgfmathsetmacro\Cxf{\Cxi+cos(-45)}
   \pgfmathsetmacro\Cyf{\Cyi+sin(-45)}
   \coordinate (pCi) at (\Cxi,\Cyi);
   \coordinate (pCf) at (\Cxf,\Cyf);
   \pgfmathsetmacro\Dxi{\cx+\r*cos(-135)}
   \pgfmathsetmacro\Dyi{\cy+\r*sin(-135)}
   \pgfmathsetmacro\Dxf{\Dxi+cos(-135)}
   \pgfmathsetmacro\Dyf{\Dyi+sin(-135)}
   \coordinate (pDi) at (\Dxi,\Dyi);
   \coordinate (pDf) at (\Dxf,\Dyf);
   \pgfmathsetmacro\Exi{\cx+\r*cos(90)}
   \pgfmathsetmacro\Eyi{\cy+\r*sin(90)}
   \pgfmathsetmacro\Exf{\Exi+cos(90)}
   \pgfmathsetmacro\Eyf{\Eyi+sin(90)}
   \coordinate (pEi) at (\Exi,\Eyi);
   \coordinate (pEf) at (\Exf,\Eyf);
   \coordinate (ti) at ($(pDf)-(.25,0)$);
   \coordinate (tf) at ($(pAf)-(.25,0)$);
   \def\rb{.55}
   \pgfmathsetmacro\sax{\cx+\rb*cos(180)}
   \pgfmathsetmacro\say{\cy+\rb*sin(180)}
   \coordinate[label=below:{\scalebox{0.5}{$x_1$}}] (s1) at (\sax,\say);
   \pgfmathsetmacro\sbx{\cx+\rb*cos(135)}
   \pgfmathsetmacro\sby{\cy+\rb*sin(135)}
   \coordinate (s2) at (\sbx,\sby);
   \pgfmathsetmacro\scx{\cx+\rb*cos(90)}
   \pgfmathsetmacro\scy{\cy+\rb*sin(90)}
   \coordinate (s3) at (\scx,\scy);
   \pgfmathsetmacro\sdx{\cx+\rb*cos(45)}
   \pgfmathsetmacro\sdy{\cy+\rb*sin(45)}
   \coordinate (s4) at (\sdx,\sdy);
   \pgfmathsetmacro\sex{\cx+\rb*cos(0)}
   \pgfmathsetmacro\sey{\cy+\rb*sin(0)}
   \coordinate[label=below:{\scalebox{0.5}{$x_3$}}] (s5) at (\sex,\sey);
   \coordinate[label=below:{\scalebox{0.5}{$x_2$}}] (sc) at (\cx,\cy);
   \draw [fill] (s1) circle (1pt);
   \draw [fill] (sc) circle (1pt);
   \draw [fill] (s5) circle (1pt);
   \draw[-,thick] (s1) -- (sc) -- (s5);
  \end{scope}
 \end{tikzpicture}
 \caption{Feynman graph contribution to the wavefunction (left) and related reduced graph (right). The reduced graphs are obtained from the Feynman ones by suppressing the external edges.}
 \label{Fig:G}
\end{figure*}

At loop level the facet is more interesting. We will show that any further face of the scattering facet corresponds to a choice of connected subgraph, and factorizes into a product of lower-dimensional scattering facets, together with a simplex. This will turn out to directly imply the unitarity of scattering amplitudes.

Lorentz invariance is more striking: the wavefunction and the polytope know only about spatial momenta, so how can we see manifest Lorentz invariance at loop level, where the ``$l_0$'' parts of the integration are also needed? The answer is simple and beautiful. The canonical form for any polytope $\mathcal{P}$, given by the convex hull of a collection of vertices ${\bf V}_j$, has a contour integral representation of the form \cite{Arkani-Hamed:2017tmz}:
\begin{equation*}
 \Omega(\mathcal{P, {\mathcal Y}})\:=\:\int\prod_{j}\frac{dc_j}{c_j-i\varepsilon_j}\delta(\mathcal{Y} - c_j {\bf V}_j),
\end{equation*}
This representation is closely related to the fact that the canonical form of a polytope ${\cal P}$ is given by the volume of the dual polytope $\tilde{\mathcal P}$ relative to ${\mathcal Y}$ as the hyperplane at infinity. 

When applied to the cosmological polytope, and on the support of the delta functions, the remaining $c_j$ integrals precisely turn into the $l_0$ integrations making Lorentz invariance manifest, with the correct Feynman $i\varepsilon$ forced by the standard prescription for the canonical form.

Thus the scattering facet of the cosmological polytope gives us a combinatorial origin for exact Lorentz invariance and unitarity, without having these as primary concepts. It is quite pleasing that unitarity is made obvious by the polytope itself, while Lorentz invariance is most naturally understood by thinking in terms of the dual polytope. This understanding of the scattering facet also gives us a conceptually transparent understanding of the cutting rules for scattering amplitudes of general quantum field theories, distinct from the clever arguments making use of the largest time equation \cite{Veltman:1994wz}.


\section{Cosmological polytope} \label{sec:WFCP}

We begin with the path-integral representation for the wavefunction of a class of scalar field theories in $d+1$ dimensions
\begin{equation}
 \psi[\phi]\:=\:\int \mathcal{D}\varphi\,e^{iS[\phi_0+\varphi]}
\end{equation}
where $\phi_0\,=\,\phi(p)\,e^{iE_p\eta}$ is the free solution with oscillatory behavior ensuring the Bunch-Davies vacuum in the infinite past, while the fluctuations $\varphi$ vanish at the time $\eta_0$ associated with the future spatial slice on which the wavefunction is computed. In what follows, $S[\phi]$ is taken to be the action for a scalar in flat space with time-dependent polynomial interactions:
\begin{equation*}
 S[\phi] = \int d^d x d \eta\left[\frac{1}{2}\left(\partial\phi\right)^2 - \sum_{k\ge3}\frac{\lambda_k(\eta)}{k!}\phi^k\right].
\end{equation*}
This class of theories includes as a special example the case of conformally-coupled scalars with non-conformal polynomial interactions in FRW cosmologies, where the $\lambda_k(\eta) = \left[a(\eta)\right]^{(2-k)(d-1)/2+2}$. 

Perturbatively, the wavefunction can be computed via Feynman graphs. It is convenient to Fourier represent the couplings: $\lambda_k(\eta) = \int dE e^{i E \eta} \tilde{\lambda}(E)$.
We then solve for the wavefunction with this oscillatory time dependence, leaving the integrals over $E$ to the end. This leads us to focus on an ``integrand" for the perturbative contributions to the wavefunction, which is a rational function of energy variables (for more details see \cite{Arkani-Hamed:2017fdk}). For this simple scalar theory with non-derivative interactions, the dependence on the energy further simplifies, depending on the sum $x_i\,\equiv\,\sum_{k\in v_i}E_k$ of the energies of the external states $E_k\,\equiv\,|\overrightarrow{p}_k|$ at each vertex $v_i$, and on the internal energies $y_{ij}$ associated with the edges between the vertices $v_i$ and $v_j$. Note of course that there the internal energies are given by the magnitude of the internal spatial momenta. When the graph has loops, we must also integrate as usual over the undetermined loop momentum, but of course here we integrate only spatial loop momenta $\mathbf{\ell}$.

Thus associated with every Feynman graph ${\cal G}$, we have (the integrand of) its corresponding contribution to the perturbative wavefunction, $\Psi_{\cal{G}}(x_v,y_e)$. Concretely, this representation of $\Psi_{\cal G}(x_v,y_e)$ is given by a time-integral: 
\begin{equation}
\Psi_{\cal {G}}(x_v,y_e) = \int_{-\infty}^0 \prod_{v\in\mathcal{V}} d \eta_v e^{ i x_v \eta_v} \prod_{e\in\mathcal{E}} G(\eta_v, \eta_{v^\prime},y_e)
\end{equation}
where $\mathcal{V}$ and $\mathcal{E}$ are the sets of vertices and edges of $\mathcal{G}$ respectively, and
$(2y)G(\eta, \eta^\prime, y) = \Theta(\eta - \eta^\prime) e^{i y (\eta^\prime - \eta)} + \Theta(\eta^\prime - \eta) e^{i y (\eta - \eta^\prime)} - e^{i y (\eta + \eta^\prime)}$
 is the bulk-bulk propagator appropriate to the computation of the wavefunction. 

But as pointed out in \cite{Arkani-Hamed:2017fdk}, $\Psi_{\cal{G}}(x_v,y_e)$ is also the answer to a completely different, natural mathematical question, and is determined by the canonical form of the {\it cosmological polytope} associated with ${\cal G}$.

Let us review the intrinsic definition of these polytopes. Consider the space of $n_e$ triangles $\triangle_i$, whose midpoints are identified by the vectors $({\bf x}_i,{\bf x'}_i, {\bf y}_i)$ -- this uniquely fixes its vertices to be $\left\{{\bf x}_i+{\bf x'}_i-{\bf y}_i,\,{\bf x}_i-{\bf x'}_i+{\bf y}_i,\,-{\bf x}_i+{\bf x'}_i+{\bf y}_i\right\}$. Each $\triangle_i$ is characterized by two edges on which it can intersect any other triangle $\triangle_j$ identifying the respective midpoints, and the third edge on which no intersection is allowed. Examples with two intersecting triangles are shown below; in the first, the two triangles intersect on one edge, in the second they intersect on both:

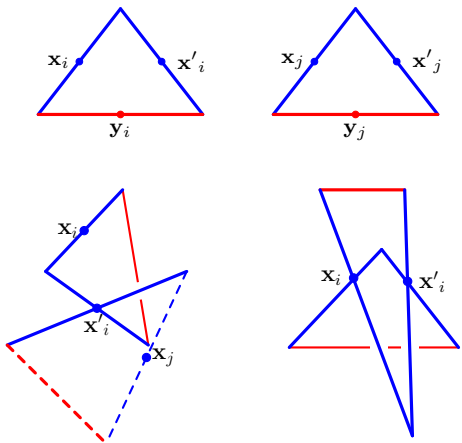
\begin{equation*}
 \begin{tikzpicture}[line join = round, line cap = round, ball/.style = {circle, draw, align=center, anchor=north, inner sep=0},
                     axis/.style={very thick, ->, >=stealth'}, pile/.style={thick, ->, >=stealth', shorten <=2pt, shorten>=2pt}, every node/.style={color=black}, scale={1.25}]
  \begin{scope}[shift={(4,2.5)}, scale={.5}]
   \coordinate (A) at (0,0);
   \coordinate (B) at (-1.75,-2.25);
   \coordinate (C) at (+1.75,-2.25);
   \coordinate [label=left:{\footnotesize $\displaystyle {\bf x}_i$}] (m1) at ($(A)!0.5!(B)$);
   \coordinate [label=right:{\footnotesize $\displaystyle \;{\bf x'}_i$}] (m2) at ($(A)!0.5!(C)$);
   \coordinate [label=below:{\footnotesize $\displaystyle {\bf y}_i$}] (m3) at ($(B)!0.5!(C)$);
   \tikzset{point/.style={insert path={ node[scale=2.5*sqrt(\pgflinewidth)]{.} }}}

   \draw[color=blue,fill=blue] (m1) circle (2pt);
   \draw[color=blue,fill=blue] (m2) circle (2pt);
   \draw[color=red,fill=red] (m3) circle (2pt);

   \draw[-, very thick, color=blue] (B) -- (A);
   \draw[-, very thick, color=blue] (A) -- (C);
   \draw[-, very thick, color=red] (B) -- (C);
  \end{scope}
  \begin{scope}[shift={(6.5,2.5)}, scale={.5}]
   \coordinate (A) at (0,0);
   \coordinate (B) at (-1.75,-2.25);
   \coordinate (C) at (+1.75,-2.25);
   \coordinate [label=left:{\footnotesize $\displaystyle {\bf x}_j$}] (m1) at ($(A)!0.5!(B)$);
   \coordinate [label=right:{\footnotesize $\displaystyle \;{\bf x'}_j$}] (m2) at ($(A)!0.5!(C)$);
   \coordinate [label=below:{\footnotesize $\displaystyle {\bf y}_j$}] (m3) at ($(B)!0.5!(C)$);
   \tikzset{point/.style={insert path={ node[scale=2.5*sqrt(\pgflinewidth)]{.} }}}

   \draw[color=blue,fill=blue] (m1) circle (2pt);
   \draw[color=blue,fill=blue] (m2) circle (2pt);
   \draw[color=red,fill=red] (m3) circle (2pt);

   \draw[-, very thick, color=blue] (B) -- (A);
   \draw[-, very thick, color=blue] (A) -- (C);
   \draw[-, very thick, color=red] (B) -- (C);
  \end{scope}
  \begin{scope}[scale={.5}, shift={(7.5,2)}, transform shape]
   \pgfmathsetmacro{\factor}{1/sqrt(2)};
   \coordinate  (B2) at (1.5,-3,-1.5*\factor);
   \coordinate  (A1) at (-1.5,-3,-1.5*\factor);
   \coordinate  (B1) at (1.5,-3.75,1.5*\factor);
   \coordinate  (A2) at (-1.5,-3.75,1.5*\factor);
   \coordinate  (C1) at (0.75,-.65,.75*\factor);
   \coordinate  (C2) at (0.4,-6.05,.75*\factor);
   \coordinate (Int) at (intersection of A2--B2 and B1--C1);
   \coordinate (Int2) at (intersection of A1--B1 and A2--B2);

   \tikzstyle{interrupt}=[
    postaction={
        decorate,
        decoration={markings,
                    mark= at position 0.5
                          with
                          {
                            \node[rectangle, color=white, fill=white, below=-.1 of Int] {};
                          }}}
   ]

   \draw[interrupt,thick,color=red] (B1) -- (C1);
   \draw[-,very thick,color=blue] (A1) -- (B1);
   \draw[-,very thick,color=blue] (A2) -- (B2);
   \draw[-,very thick,color=blue] (A1) -- (C1);
   \draw[-, dashed, very thick, color=red] (A2) -- (C2);
   \draw[-, dashed, thick, color=blue] (B2) -- (C2);

   \coordinate[label=below:{\Large ${\bf x'}_i$}] (x2) at ($(A1)!0.5!(B1)$);
   \draw[fill,color=blue] (x2) circle (2.5pt);
   \coordinate[label=left:{\Large ${\bf x}_i$}] (x1) at ($(C1)!0.5!(A1)$);
   \draw[fill,color=blue] (x1) circle (2.5pt);
   \coordinate[label=right:{\Large ${\bf x}_j$}] (x3) at ($(B2)!0.5!(C2)$);
   \draw[fill,color=blue] (x3) circle (2.5pt);
  \end{scope}
  \begin{scope}[scale={.6}, shift={(10.75,.75)}, transform shape]
   \pgfmathsetmacro{\factor}{1/sqrt(2)};
   \coordinate  (c1b) at (0.75,0,-.75*\factor);
   \coordinate  (b1b) at (-.75,0,-.75*\factor);
   \coordinate  (a2b) at (0.75,-.65,.75*\factor);

   \coordinate  (c2b) at (1.5,-3,-1.5*\factor);
   \coordinate  (b2b) at (-1.5,-3,-1.5*\factor);
   \coordinate  (a1b) at (1.5,-3.75,1.5*\factor);

   \coordinate (Int1) at (intersection of b2b--c2b and b1b--a1b);
   \coordinate (Int2) at (intersection of b2b--c2b and c1b--a1b);
   \coordinate (Int3) at (intersection of b2b--a2b and b1b--a1b);
   \coordinate (Int4) at (intersection of a2b--c2b and c1b--a1b);
   \tikzstyle{interrupt}=[
    postaction={
        decorate,
        decoration={markings,
                    mark= at position 0.5
                          with
                          {
                            \node[rectangle, color=white, fill=white] at (Int1) {};
                            \node[rectangle, color=white, fill=white] at (Int2) {};
                          }}}
   ]

   \node at (c1b) (c1c) {};
   \node at (b1b) (b1c) {};
   \node at (a2b) (a2c) {};
   \node at (c2b) (c2c) {};
   \node at (b2b) (b2c) {};
   \node at (a1b) (a1c) {};

   \draw[interrupt,thick,color=red] (b2b) -- (c2b);
   \draw[-,very thick,color=red] (b1b) -- (c1b);
   \draw[-,very thick,color=blue] (b1b) -- (a1b);
   \draw[-,very thick,color=blue] (a1b) -- (c1b);
   \draw[-,very thick,color=blue] (b2b) -- (a2b);
   \draw[-,very thick,color=blue] (a2b) -- (c2b);

   \node[ball,text width=.15cm,fill,color=blue, above=-.06cm of Int3, label=left:{\large ${\bf x}_i$}] (Inta) {};
   \node[ball,text width=.15cm,fill,color=blue, above=-.06cm of Int4, label=right:{\large ${\bf x'}_i$}] (Intb) {};

  \end{scope}
 \end{tikzpicture}
\end{equation*}

Note that any such collection of intersecting triangles is naturally associated with a graph. Every triangle is represented by an edge ending on two vertices; these vertices represent the two sides on which the triangle can intersect other triangles.
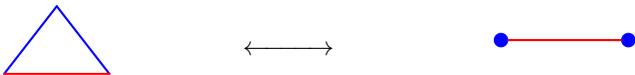
\begin{equation*}
 \begin{tikzpicture}[shift={(1,0)}, line join = round, line cap = round, ball/.style = {circle, draw, align=center, anchor=north, inner sep=0}, scale={.4}]
  \coordinate (A) at (0,0);
  \coordinate (B) at (-1.75,-2.25);
  \coordinate (C) at (+1.75,-2.25);
  \coordinate  (m2) at ($(A)!0.5!(C)$);

  \draw[-, thick, color=blue] (B) -- (A); 
  \draw[-, thick, color=blue] (A) -- (C); 
  \draw[-, thick, color=red] (B) -- (C);

   \node[right=2cm of m2.east] (lra) {$\xleftrightarrow{\phantom{convext}}$};
   \node[ball,text width=.18cm,fill,color=blue,right=2cm of lra.east] (x1) {};
    \node[ball,text width=.18cm,fill,color=blue,right=1.5cm of x1.east] (x2) {};
    \draw[-,thick,color=red] (x1.east) -- (x2.west);
   \end{tikzpicture}
\end{equation*}
If two triangles do intersect on a midpoint, their corresponding vertices are joined, producing a graph associated with this collection of intersecting triangles; some examples are shown below:
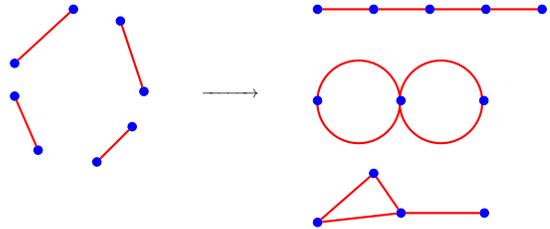
\begin{equation*}
 \begin{tikzpicture}[ball/.style = {circle, draw, align=center, anchor=north, inner sep=0}, scale={.625}, transform shape]
  \node[ball,text width=.18cm,fill,color=blue] at (-.5,0) (x1) {};
  \node[ball,text width=.18cm,fill,color=blue] at (-1.75,-1.15) (x2) {};
  \node[ball,text width=.18cm,fill,color=blue] at (-1.75,-1.85) (x3) {};
  \node[ball,text width=.18cm,fill,color=blue] at (.5,-.25) (x4) {};
  \node[ball,text width=.18cm,fill,color=blue] at (1,-1.75) (x5) {};
  \node[ball,text width=.18cm,fill,color=blue] at (-1.25,-3) (x6) {};
  \node[ball,text width=.18cm,fill,color=blue] at (.75,-2.5) (x7) {};
  \node[ball,text width=.18cm,fill,color=blue] at (0,-3.25) (x8) {};

  \draw[-,thick,color=red] (x1) -- (x2);
  \draw[-,thick,color=red] (x4) -- (x5);
  \draw[-,thick,color=red] (x3) -- (x6);
  \draw[-,thick,color=red] (x7) -- (x8);

  \node[right=1cm of x5.east] (arr) {$\displaystyle \xrightarrow{\hspace{1cm}}$};

  \node[ball,text width=.18cm,fill,color=blue,right=5cm of x1.east] (xa) {};
  \node[ball,text width=.18cm,fill,color=blue,right=1cm of xa.east] (xb) {};
  \node[ball,text width=.18cm,fill,color=blue,right=1cm of xb.east] (xc) {};
  \node[ball,text width=.18cm,fill,color=blue,right=1cm of xc.east] (xd) {};
  \node[ball,text width=.18cm,fill,color=blue,right=1cm of xd.east] (xe) {};
  \draw[-,thick,color=red] (xa.east) -- (xb.west);
  \draw[-,thick,color=red] (xb.east) -- (xc.west);
  \draw[-,thick,color=red] (xc.east) -- (xd.west);
  \draw[-,thick,color=red] (xd.east) -- (xe.west);

  \node[below=1.75cm of xa] (t1) {};
  \node[right=1.5cm of t1] (t2) {};
  \node[right=1.5cm of t2] (t3) {};

   \draw[thick, color=red] ($(t1)!0.5!(t2)$) circle (.88);
   \draw[thick, color=red] ($(t2)!0.5!(t3)$) circle (.88);

  \node[ball,text width=.18cm,fill,color=blue,below=1.75cm of xa] (xf) {};
  \node[ball,text width=.18cm,fill,color=blue,right=1.58cm of xf.east] (xg) {};
  \node[ball,text width=.18cm,fill,color=blue,right=1.57cm of xg.east] (xh) {};

  \node[ball,text width=.18cm,fill,color=blue,below=3.3cm of xb] (xi) {};
  \node[ball,text width=.18cm,fill,color=blue,below=2.4cm of xf] (xj) {};
  \node[ball,text width=.18cm,fill,color=blue,below=2.2cm of xg] (xk) {};
  \node[ball,text width=.18cm,fill,color=blue,right=1.58cm of xk.east] (xl) {};

  \draw[-,thick,color=red] (xi) -- (xj) -- (xk) -- (xi);
  \draw[-,thick,color=red] (xk) -- (xl);

 \end{tikzpicture}
\end{equation*}

The cosmological polytope $\mathcal{P}$ is the convex hull of the $3n_e$ vertices of $n_e$ intersected triangles. Very concretely, starting with the graph ${\cal G}$ rather than the picture of intersecting triangles: for any ${\cal G}$, we can associated vectors ${\bf x}_v$ with all the vertices and ${\bf y}_e$ with all the edges. These vectors taken together give a basis for a projective space $\mathbf{P}^{n_e + n_v - 1}$. Each edge of the ${\cal G}$ is associated with the three vertices as above: $\left\{{\bf x}_i+{\bf x'}_i-{\bf y}_i,\,{\bf x}_i-{\bf x'}_i+{\bf y}_i,\,-{\bf x}_i+{\bf x'}_i+{\bf y}_i\right\}$. The cosmological polytope is the convex hull of these $3 n_e$ vertices.

The definition of the cosmological polytope is extremely simple; its only unusual feature is the asymmetry between the sides of the triangle. What is the reason for the ``three" associated with using triangles, and why are two of the three sides distinguished? As discussed in \cite{Arkani-Hamed:2017fdk} and as we will see in further action in this letter, these features are crucial: they are the primitive combinatorial avatar of causal space-time structure. Indeed we know that lightcones emanating from a point in space-time divides space-time into three regions, with two of these (past and future time-like separated) of a different type than the third (space-like separated). This crucial physics is captured by the primitive intersecting triangle rules.

If we write any point of $\mathcal{P}$ as $\mathcal{Y}\,=\,\sum_v x_v{\bf X}_v+\sum_e y_e{\bf Y}_e$, where ${\bf X}_v$ and ${\bf Y}_e$ are vectors in $\mathbb{R}^{n_e+n_v}$ identifying the independent midpoints ${\bf x}$ and ${\bf y}$ of the triangles generating $\mathcal{P}$, then the coefficients $x_v$ and $y_e$ will label the vertices and the edges of $\mathcal{G}$ respectively.

Now, {\it any} polytope $\mathcal{P}$ in a projective space $\mathbb{P}^N$ with co-ordinates ${\cal Y}$ has an associated canonical differential top form $\omega({\cal Y}; {\cal P})$, uniquely fixed by the property of having logarithmic singularities on (and only on) all faces of all dimensionality of ${\cal P}$. It is convenient to also associate a function $\Omega({\cal Y}, {\cal P})$ by pulling out a universal top-form measure on the projective space as $\omega({\cal Y}, {\cal{P}}) = \langle\mathcal{Y} d^N \mathcal{Y}\rangle\,\Omega({\cal Y},{\cal P})$. For the cosmological polytope, this canonical function directly determines the (integrand of) the wavefunction $\Psi_{\mathcal{G}}(x_v,y_e)$ for the graph $\mathcal{G}$ \cite{Arkani-Hamed:2017fdk}:
\begin{equation}\label{eq:CF}
 \Omega(\mathcal{Y};\,\mathcal{P})\:=\: \Psi_{\mathcal{G}}(x_v,y_e)
\end{equation}
and any graph $\mathcal{G}$ is nothing but a Feynman graph with the external edges suppressed (see Figure \ref{Fig:G}). In some cases, such as tree diagrams for $\phi^3$ theory in $dS_4$, the integrand $\Psi$ can be integrated to yield the final contribution of the graph to the wavefunction, and these functions turn out to be interesting polylogarithms. As described in \cite{Arkani-Hamed:2017fdk}, the ``symbol" of these polylogs is also directly determined by the geometry of the cosmological polytope, but in this letter we will focus on the properties of the integrand, which are all what we need to see the emergence of Lorentz invariance and unitarity.

Our definition of the cosmological polytope was given as a convex hull of a collection of vertices, but its beautiful combinatorial structure allows us to completely characterize all of its facets as well. This is important since the canonical forms on the facets compute the residues of the wavefunction on its poles. 

As shown in \cite{Arkani-Hamed:2017fdk}, the (codimension one) facets can be found as hyperplanes $\mathcal{W}_I\,=\,\tilde{x}_v{\bf \tilde{X}}_{vI}+\tilde{y}_e{\bf \tilde{Y}}_{eI}$ 
(where $(\tilde{\bf {X}}_v\cdot {\bf x}_{v'})=\delta_{vv'}$, $\tilde{{\bf y}}_e\cdot {\bf y}_{e'}=\delta_{ee'}$, $(\tilde{ \bf{x}}_v\cdot {\bf y}_e)=0$) such that, given the collection of vertices ${\bf V}_a^I$ ($a\,=\,1,\,\ldots,\,3n_e$) of $\mathcal{P}$, $\mathcal{W}_I{\bf V}_a^I\,\ge\,0$ with the maximum number of $\mathcal{W}_I{\bf V}_a^I$ set to zero, compatible with the constraints on the midpoints. This is equivalent to finding a pattern for $\alpha_{(e,e)}\equiv\tilde{x}_v+\tilde{x}_{v'}-\tilde{y}_e\ge0$, $\alpha_{(e,v)}\equiv\tilde{x}_v+\tilde{y}_e-\tilde{x}_{v'}\ge0$, $\alpha_{(e,v')}\equiv\tilde{x}_{v'}+\tilde{y}_e-\tilde{x}_v\ge0$, with at least one non-zero $\alpha$, such that setting any other $\alpha$ to zero the compatibility relation with the midpoint constraints, now expressible as $\alpha_{(e,e)}+\alpha_{(e,v)}=\alpha_{(e',e')}+\alpha_{(e',v)}$ for any two edges $e$ and $e'$ with a common vertex $v$, forces all the other $\alpha$'s to vanish. Graphically, this can be done by marking the edges of a given graph $\mathcal{G}$ according to the non-vanishing $\alpha$'s:
\begin{equation*}
 \begin{tikzpicture}[ball/.style = {circle, draw, align=center, anchor=north, inner sep=0}, cross/.style={cross out, draw, minimum size=2*(#1-\pgflinewidth), inner sep=0pt, outer sep=0pt}, scale={1.125}, transform shape]
  \begin{scope}
   \node[ball,text width=.18cm,fill,color=black,label=below:{\footnotesize $v\phantom{'}$}] at (0,0) (v1) {};
   \node[ball,text width=.18cm,fill,color=black,label=below:{\footnotesize $v'$},right=1.5cm of v1.east] (v2) {};
   \draw[-,thick,color=black] (v1.east) edge node [text width=.18cm,below=.1cm,midway] {\footnotesize $e$} (v2.west);
   \node[very thick, cross=4pt, rotate=0, color=blue, right=.7cm of v1.east]{};
   \coordinate (x) at ($(v1)!0.5!(v2)$);
   \node[right=1.25cm of v2, scale=.8] (lb1) {$\alpha_{\mbox{\tiny $(e,e)$}}\,=\,\mathcal{W}\cdot({\bf x}_v+{\bf x}_{v'}-{\bf y}_e)>\,0$};
  \end{scope}
  \begin{scope}[shift={(0,-1)}]
   \node[ball,text width=.18cm,fill,color=black,label=below:{\footnotesize $v\phantom{'}$}] at (0,0) (v1) {};
   \node[ball,text width=.18cm,fill,color=black,label=below:{\footnotesize $v'$},right=1.5cm of v1.east] (v2) {};
   \draw[-,thick,color=black] (v1.east) edge node [text width=.18cm,below=.1cm,midway] {\footnotesize $e$} (v2.west);
   \node[very thick, cross=4pt, rotate=0, color=blue, left=.1cm of v2.west]{};
   \coordinate (x) at ($(v1)!0.5!(v2)$);
   \node[right=1.25cm of v2, scale=.8] (lb1) {$\alpha_{\mbox{\tiny $(e,v')$}}\,=\,\mathcal{W}\cdot({\bf x}_{v'}+{\bf y}_e-{\bf x}_v)>\,0$};
  \end{scope}
  \begin{scope}[shift={(0,-2)}]
   \node[ball,text width=.18cm,fill,color=black,label=below:{\footnotesize $v\phantom{'}$}] at (0,0) (v1) {};
   \node[ball,text width=.18cm,fill,color=black,label=below:{\footnotesize $v'$},right=1.5cm of v1.east] (v2) {};
   \draw[-,thick,color=black] (v1.east) edge node [text width=.18cm,below=.1cm,midway] {\footnotesize $e$} (v2.west);
   \node[very thick, cross=4pt, rotate=0, color=blue, right=.1cm of v1.east]{};
   \coordinate (x) at ($(v1)!0.5!(v2)$);
   \node[right=1.25cm of v2, scale=.8] (lb1) {$\alpha_{\mbox{\tiny $(e,v)$}}\,=\,\mathcal{W}\cdot({\bf x}_v+{\bf y}_e-{\bf x}_{v'})>\,0$};
  \end{scope}
 \end{tikzpicture}
\end{equation*}
The vertices related to the marking {\it do not} lie on $\mathcal{W}$. Thus, given a graph $\mathcal{G}$, any facet of the related cosmological polytope $\mathcal{P}$ can be identified by the consistent markings of $\mathcal{G}$, and its vertices are the ones for which the $\alpha$'s are zero, {\it i.e.} all except the ones identified by the marking. The consistent markings can be identified by considering any subgraph $\mathfrak{g}$ of $\mathcal{G}$ and marking the edges internal to $\mathfrak{g}$ in their middle, while the external edges ending on vertices $v$ of $\mathfrak{g}$ next to $v$. As a special case, let us consider $\mathfrak{g}=\mathcal{G}$:
\begin{equation*}
 \begin{tikzpicture}[ball/.style = {circle, draw, align=center, anchor=north, inner sep=0}, cross/.style={cross out, draw, minimum size=2*(#1-\pgflinewidth), inner sep=0pt, outer sep=0pt}, scale={.9}, transform shape]
  \node[ball,text width=.18cm,fill,color=black,label=below:{$v_1$}] at (0,0) (v1) {};
  \node[ball,text width=.18cm,fill,color=black,right=1.5cm of v1.east] (v2) {};
  \node[ball,text width=.18cm,fill,color=black,right=1.5cm of v2.east] (v3) {};
  \node[ball,text width=.18cm,fill,color=black,right=1.5cm of v3.east] (v4) {};
  \node[ball,text width=.18cm,fill,color=black,above=1cm of v4.north] (v5) {};
  \node[ball,text width=.18cm,fill,color=black,above=1cm of v5.north] (v6) {};
  \draw[-,thick,color=black] (v1) edge  (v2);
  \draw[-,thick,color=black] (v1) edge node[very thick,cross=4pt,rotate=0,color=blue,midway] {} (v2);
  \draw[-,thick,color=black] (v2) edge  (v3);
  \draw[-,thick,color=black] (v2) edge node[very thick,cross=4pt,rotate=0,color=blue,midway] {} (v3);
  \draw[-,thick,color=black] (v3) edge (v4);
  \draw[-,thick,color=black] (v3) edge node[very thick,cross=4pt,rotate=0,color=blue,midway] {} (v4);
  \draw[-,thick,color=black] (v4) edge node[very thick,cross=4pt,rotate=0,color=blue,midway] {} (v5);
  \draw[thick,color=black] ($(v2)!0.5!(v3)$) circle (.85cm);
  \draw[thick,color=black] ($(v5)!0.5!(v6)$) circle (.6cm);

  \coordinate (t1) at ($(v2)!0.5!(v3)$);
  \node[very thick,cross=4pt,rotate=0,color=blue,above=.76cm of t1.north] {};
  \node[very thick,cross=4pt,rotate=0,color=blue,below=.76cm of t1.south] {};

  \coordinate (t2) at ($(v5)!0.5!(v6)$);
  \node[very thick,cross=4pt,rotate=0,color=blue,left=.5cm of t2.west] {};
  \node[very thick,cross=4pt,rotate=0,color=blue,right=.5cm of t2.east] {};
 \end{tikzpicture}
\end{equation*}
all the edges are marked on their middles, {\it i.e.} all the $\alpha_{(e,e')}$ are non-zero. Such a facet is the polytope identified by the collection of $2n_e$ vertices $\left\{{\bf x}_v+{\bf y}_{e}-{\bf x}_{v'},\,-{\bf x}_v+{\bf y}_{e}+{\bf x}_{v'}\right\}$. It is easy to check that the hyperplane $\mathcal{W}$ such that $\mathcal{W}\cdot{\bf V}$ for any vertex ${\bf V}$ belonging to such a collection, is given by $\mathcal{W}=\sum_v\tilde{x}_v{\bf\tilde{X}}_v$, {\it i.e.} this facet is identified by the total energy vanishing and, thus, encodes the flat-space amplitude. This is the ``scattering facet" we will focus on in the rest of this letter. 
\begin{figure}
 \centering
 \begin{tikzpicture}[line join = round, line cap = round, ball/.style = {circle, draw, align=center, anchor=north, inner sep=0}]
  \begin{scope}[scale={.85}, transform shape]
   \pgfmathsetmacro{\factor}{1/sqrt(2)};
   \coordinate[label=above:{${\bf 2}$}]  (c1a) at (9.75,0,-.75*\factor);
   \coordinate[label=above:{${\bf 1}$}]   (b1a) at (8.25,0,-.75*\factor);
   \coordinate[label=below:{${\bf 4}$}]   (c2a) at (10.5,-3,-1.5*\factor);
   \coordinate[label=below:{${\bf 3}$}]   (b2a) at (7.5,-3,-1.5*\factor);

   \draw[-,fill=red,opacity=.5,thick] (c1a) -- (b1a) -- (b2a) -- (c2a) -- cycle;
  \end{scope}

  \begin{scope}[shift={(11,-2.5)}, scale={.75}, transform shape]
   \def\r{2.5}
   \pgfmathsetmacro\ax{\r*cos(30)}
   \pgfmathsetmacro\ay{\r*sin(30)}
   \coordinate[label=left:{${\bf 3}$}] (v3) at (0,0);
   \coordinate[label=above:{${\bf 1}$}] (v1) at ($(v3)+(0,3)$);
   \coordinate[label=below:{${\bf 5}$}] (v5) at ($(v3)+(3,0)$);
   \coordinate[label=above:{${\bf 4}$}] (v4) at ($(v3)+(\ax,\ay)$);
   \coordinate[label=above:{${\bf 2}$}] (v2) at ($(v1)+(\ax,\ay)$);
   \coordinate[label=above:{${\bf 6}$}] (v6) at ($(v5)+(\ax,\ay)$);

   \draw[-,dashed,fill=red!40, opacity=.6] (v3) -- (v4) -- (v6) -- (v5) -- cycle;
   \draw[-,dashed,fill=blue!40, opacity=.6] (v3) -- (v4) -- (v2) -- (v1) -- cycle;
   \draw[-,dashed,fill=red!60, opacity=.2] (v6) -- (v4) -- (v2) -- cycle;
   \draw[-,thick, fill=blue!60, opacity=.2] (v3) -- (v5) -- (v1) -- cycle;
   \draw[-,thick, fill=yellow!60!black, opacity=.5] (v1) -- (v2) -- (v6) -- (v5) -- cycle;
  \end{scope}
 \end{tikzpicture}
 \caption{Scattering facets for the one-loop (left) and two-loop (right) two-site graphs. They are identified by the vertices $\{{\bf 1},\,{\bf 2},\,{\bf 3},\,{\bf 4}\}\,\equiv\,\{{\bf x}_1+{\bf y}_a-{\bf x}_2,\,-{\bf x}_1+{\bf y}_a+{\bf x}_2\,{\bf x}_1+{\bf y}_b-{\bf x}_{2},\,-{\bf x}_1+{\bf y}_b+{\bf x}_2\}$ and $\{{\bf 1},\,{\bf 2},\,{\bf 3},\,{\bf 4},\,{\bf 5},\,{\bf 6}\}\,\equiv\,\{{\bf x}_1+{\bf y}_a-{\bf x}_2,\,-{\bf x}_1+{\bf y}_a+{\bf x}_2,\,{\bf x}_1+{\bf y}_b-{\bf x}_2,\,-{\bf x}_1+{\bf y}_b+{\bf x}_2,\,{\bf x}_1+{\bf y}_c-{\bf x}_2,\,-{\bf x}_1+{\bf y}_c+{\bf x}_2\}$ respectively.}
\end{figure}


\section{Emergent unitarity}\label{sec:EUn}

The scattering facet is itself a polytope with an interesting face structure. We will now see that the faces of the scattering facet have the geometric structure of the direct product of smaller scattering facets, together with a certain simplex. The residue of the canonical form factorizes in the same way. This factorization can then be precisely interpreted as the usual cutting rules for the $S$-matrix; with the extra simplex accounting for the Lorentz-invariant phase space of cut internal lines.

The lower-dimensional faces of the scattering facet are given by the collection of vertices that are on the scattering face and correspond to  a subgraph. Given a graph $\mathcal{G}$, its scattering facet is given by marking all the edges in the middle, and its vertices are the same of the cosmological polytope but the ones related to the marking. Let us now take a subgraph $\mathfrak{g}$ of $\mathcal{G}$, to which a further marking is associated, as explained earlier:
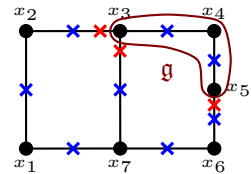
\begin{equation*}
 \begin{tikzpicture}[ball/.style = {circle, draw, align=center, anchor=north, inner sep=0}, cross/.style={cross out, draw, minimum size=2*(#1-\pgflinewidth), inner sep=0pt, outer sep=0pt}, scale=1.25, transform shape]
  \begin{scope}
   \coordinate[label=below:{\tiny $x_1$}] (v1) at (0,0);
   \coordinate[label=above:{\tiny $x_2$}] (v2) at ($(v1)+(0,1.25)$);
   \coordinate[label=above:{\tiny $x_3$}] (v3) at ($(v2)+(1,0)$);
   \coordinate[label=above:{\tiny $x_4$}] (v4) at ($(v3)+(1,0)$);
   \coordinate[label=right:{\tiny $x_5$}] (v5) at ($(v4)-(0,.625)$);
   \coordinate[label=below:{\tiny $x_6$}] (v6) at ($(v5)-(0,.625)$);
   \coordinate[label=below:{\tiny $x_7$}] (v7) at ($(v6)-(1,0)$);
   \draw[thick] (v1) -- (v2) -- (v3) -- (v4) -- (v5) -- (v6) -- (v7) -- cycle;
   \draw[thick] (v3) -- (v7);
   \draw[fill=black] (v1) circle (2pt);
   \draw[fill=black] (v2) circle (2pt);
   \draw[fill=black] (v3) circle (2pt);
   \draw[fill=black] (v4) circle (2pt);
   \draw[fill=black] (v5) circle (2pt);
   \draw[fill=black] (v6) circle (2pt);
   \draw[fill=black] (v7) circle (2pt);
   \coordinate (v12) at ($(v1)!0.5!(v2)$);
   \coordinate (v23) at ($(v2)!0.5!(v3)$);
   \coordinate (v34) at ($(v3)!0.5!(v4)$);
   \coordinate (v45) at ($(v4)!0.5!(v5)$);
   \coordinate (v56) at ($(v5)!0.5!(v6)$);
   \coordinate (v67) at ($(v6)!0.5!(v7)$);
   \coordinate (v71) at ($(v7)!0.5!(v1)$);
   \coordinate (v37) at ($(v3)!0.5!(v7)$);
   \node[very thick, cross=4pt, rotate=0, color=blue, scale=.625] at (v12) {};
   \node[very thick, cross=4pt, rotate=0, color=blue, scale=.625] at (v23) {};
   \node[very thick, cross=4pt, rotate=0, color=blue, scale=.625] at (v34) {};
   \node[very thick, cross=4pt, rotate=0, color=blue, scale=.625] at (v45) {};
   \node[very thick, cross=4pt, rotate=0, color=blue, scale=.625] at (v56) {};
   \node[very thick, cross=4pt, rotate=0, color=blue, scale=.625] at (v67) {};
   \node[very thick, cross=4pt, rotate=0, color=blue, scale=.625] at (v71) {};
   \node[very thick, cross=4pt, rotate=0, color=blue, scale=.625] at (v37) {};
   \node[very thick, cross=4pt, rotate=0, color=red, scale=.625, left=.15cm of v3] (v3l) {};
   \node[very thick, cross=4pt, rotate=0, color=red, scale=.625, below=.15cm of v3] (v3b) {};
   \node[very thick, cross=4pt, rotate=0, color=red, scale=.625, below=.1cm of v5] (v5b){};
   \coordinate (a) at ($(v3l)!0.5!(v3)$);
   \coordinate (b) at ($(v3)+(0,.125)$);
   \coordinate (c) at ($(v34)+(0,.175)$);
   \coordinate (d) at ($(v4)+(0,.125)$);
   \coordinate (e) at ($(v4)+(.125,0)$);
   \coordinate (f) at ($(v45)+(.175,0)$);
   \coordinate (g) at ($(v5)+(.125,0)$);
   \coordinate (h) at ($(v5b)!0.5!(v5)$);
   \coordinate (i) at ($(v5)-(.125,0)$);
   \coordinate (j) at ($(v45)-(.175,0)$);
   \coordinate (k) at ($(v34)-(0,.175)$);
   \coordinate (l) at ($(v3)-(0,.125)$);
   \draw [thick, red!50!black] plot [smooth cycle] coordinates {(a) (b) (c) (d) (e) (f) (g) (h) (i) (j) (k) (l)};
   \node[below=.05cm of k, color=red!50!black] {\footnotesize $\displaystyle\mathfrak{g}$};
  \end{scope}
 \end{tikzpicture}
\end{equation*}
The two markings identify a face of the scattering facet, singling out all the vertices that do not belong to it. We can indicate such a face by marking with  open circles the vertices which belong to it:
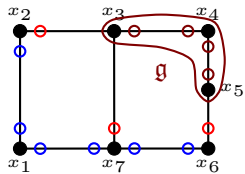
\begin{equation*}
 \begin{tikzpicture}[ball/.style = {circle, draw, align=center, anchor=north, inner sep=0}, cross/.style={cross out, draw, minimum size=2*(#1-\pgflinewidth), inner sep=0pt, outer sep=0pt}, scale=1.25, transform shape]
  \begin{scope}[shift={(0,-2)}, transform shape]
   \coordinate[label=below:{\tiny $x_1$}] (v1) at (0,0);
   \coordinate[label=above:{\tiny $x_2$}] (v2) at ($(v1)+(0,1.25)$);
   \coordinate[label=above:{\tiny $x_3$}] (v3) at ($(v2)+(1,0)$);
   \coordinate[label=above:{\tiny $x_4$}] (v4) at ($(v3)+(1,0)$);
   \coordinate[label=right:{\tiny $x_5$}] (v5) at ($(v4)-(0,.625)$);
   \coordinate[label=below:{\tiny $x_6$}] (v6) at ($(v5)-(0,.625)$);
   \coordinate[label=below:{\tiny $x_7$}] (v7) at ($(v6)-(1,0)$);
   \draw[thick] (v1) -- (v2) -- (v3) -- (v4) -- (v5) -- (v6) -- (v7) -- cycle;
   \draw[thick] (v3) -- (v7);
   \draw[fill=black] (v1) circle (2pt);
   \draw[fill=black] (v2) circle (2pt);
   \draw[fill=black] (v3) circle (2pt);
   \draw[fill=black] (v4) circle (2pt);
   \draw[fill=black] (v5) circle (2pt);
   \draw[fill=black] (v6) circle (2pt);
   \draw[fill=black] (v7) circle (2pt);
   \coordinate (v12) at ($(v1)!0.5!(v2)$);
   \coordinate (v23) at ($(v2)!0.5!(v3)$);
   \coordinate (v34) at ($(v3)!0.5!(v4)$);
   \coordinate (v45) at ($(v4)!0.5!(v5)$);
   \coordinate (v56) at ($(v5)!0.5!(v6)$);
   \coordinate (v67) at ($(v6)!0.5!(v7)$);
   \coordinate (v71) at ($(v7)!0.5!(v1)$);
   \coordinate (v37) at ($(v3)!0.5!(v7)$);
   \node[ball,text width=.18cm,thick,color=red!50!black,right=.15cm of v3, scale=.625] {};
   \node[ball,text width=.18cm,thick,color=red!50!black,left=.15cm of v4, scale=.625] {};
   \node[ball,text width=.18cm,thick,color=red!50!black,below=.1cm of v4, scale=.625] {};
   \node[ball,text width=.18cm,thick,color=red!50!black,above=.1cm of v5, scale=.625] {};
   \node[ball,text width=.18cm,thick,color=red,right=.15cm of v2, scale=.625] {};
   \node[ball,text width=.18cm,thick,color=red,above=.15cm of v7, scale=.625] {};
   \node[ball,text width=.18cm,thick,color=red,above=.15cm of v6, scale=.625] {};
   \node[ball,text width=.18cm,thick,color=blue,below=.15cm of v2, scale=.625] {};
   \node[ball,text width=.18cm,thick,color=blue,above=.15cm of v1, scale=.625] {};
   \node[ball,text width=.18cm,thick,color=blue,right=.15cm of v1, scale=.625] {};
   \node[ball,text width=.18cm,thick,color=blue,left=.15cm of v7, scale=.625] {};
   \node[ball,text width=.18cm,thick,color=blue,right=.15cm of v7, scale=.625] {};
   \node[ball,text width=.18cm,thick,color=blue,left=.15cm of v6, scale=.625] {};
   \coordinate (a) at ($(v3)-(.125,0)$);
   \coordinate (b) at ($(v3)+(0,.125)$);
   \coordinate (c) at ($(v34)+(0,.175)$);
   \coordinate (d) at ($(v4)+(0,.125)$);
   \coordinate (e) at ($(v4)+(.125,0)$);
   \coordinate (f) at ($(v45)+(.175,0)$);
   \coordinate (g) at ($(v5)+(.125,0)$);
   \coordinate (h) at ($(v5)-(0,.125)$);
   \coordinate (i) at ($(v5)-(.125,0)$);
   \coordinate (j) at ($(v45)-(.175,0)$);
   \coordinate (k) at ($(v34)-(0,.175)$);
   \coordinate (l) at ($(v3)-(0,.125)$);
   \draw [thick, red!50!black] plot [smooth cycle] coordinates {(a) (b) (c) (d) (e) (f) (g) (h) (i) (j) (k) (l)};
   \node[below=.05cm of k, color=red!50!black] {\footnotesize $\displaystyle\mathfrak{g}$};
  \end{scope}
 \end{tikzpicture}
\end{equation*}
 Furthermore, the subgraph $\mathfrak{g}$ corresponds to $\sum_{v\in\mathfrak{g}}x_v+\sum_{e\in\mathcal{E}_{\mathfrak{g}}^{\mbox{\tiny ext}}}y_e\longrightarrow0$, $\mathcal{E}_{\mathfrak{g}}^{\mbox{\tiny ext}}$ being the set of edges of $\mathcal{G}$ entering into $\mathfrak{g}$.

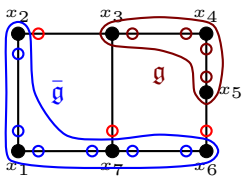
\begin{wrapfigure}{l}{3cm}
 \begin{tikzpicture}[ball/.style = {circle, draw, align=center, anchor=north, inner sep=0}, cross/.style={cross out, draw, minimum size=2*(#1-\pgflinewidth), inner sep=0pt, outer sep=0pt}, scale=1.25, transform shape]
  \begin{scope}
   \coordinate[label=below:{\tiny $x_1$}] (v1) at (0,0);
   \coordinate[label=above:{\tiny $x_2$}] (v2) at ($(v1)+(0,1.25)$);
   \coordinate[label=above:{\tiny $x_3$}] (v3) at ($(v2)+(1,0)$);
   \coordinate[label=above:{\tiny $x_4$}] (v4) at ($(v3)+(1,0)$);
   \coordinate[label=right:{\tiny $x_5$}] (v5) at ($(v4)-(0,.625)$);
   \coordinate[label=below:{\tiny $x_6$}] (v6) at ($(v5)-(0,.625)$);
   \coordinate[label=below:{\tiny $x_7$}] (v7) at ($(v6)-(1,0)$);
   \draw[thick] (v1) -- (v2) -- (v3) -- (v4) -- (v5) -- (v6) -- (v7) -- cycle;
   \draw[thick] (v3) -- (v7);
   \draw[fill=black] (v1) circle (2pt);
   \draw[fill=black] (v2) circle (2pt);
   \draw[fill=black] (v3) circle (2pt);
   \draw[fill=black] (v4) circle (2pt);
   \draw[fill=black] (v5) circle (2pt);
   \draw[fill=black] (v6) circle (2pt);
   \draw[fill=black] (v7) circle (2pt);
   \coordinate (v12) at ($(v1)!0.5!(v2)$);
   \coordinate (v23) at ($(v2)!0.5!(v3)$);
   \coordinate (v34) at ($(v3)!0.5!(v4)$);
   \coordinate (v45) at ($(v4)!0.5!(v5)$);
   \coordinate (v56) at ($(v5)!0.5!(v6)$);
   \coordinate (v67) at ($(v6)!0.5!(v7)$);
   \coordinate (v71) at ($(v7)!0.5!(v1)$);
   \coordinate (v37) at ($(v3)!0.5!(v7)$);
   \node[ball,text width=.18cm,thick,color=red!50!black,right=.15cm of v3, scale=.625] {};
   \node[ball,text width=.18cm,thick,color=red!50!black,left=.15cm of v4, scale=.625] {};
   \node[ball,text width=.18cm,thick,color=red!50!black,below=.1cm of v4, scale=.625] {};
   \node[ball,text width=.18cm,thick,color=red!50!black,above=.1cm of v5, scale=.625] {};
   \node[ball,text width=.18cm,thick,color=red,right=.15cm of v2, scale=.625] {};
   \node[ball,text width=.18cm,thick,color=red,above=.15cm of v7, scale=.625] {};
   \node[ball,text width=.18cm,thick,color=red,above=.15cm of v6, scale=.625] {};
   \node[ball,text width=.18cm,thick,color=blue,below=.15cm of v2, scale=.625] {};
   \node[ball,text width=.18cm,thick,color=blue,above=.15cm of v1, scale=.625] {};
   \node[ball,text width=.18cm,thick,color=blue,right=.15cm of v1, scale=.625] {};
   \node[ball,text width=.18cm,thick,color=blue,left=.15cm of v7, scale=.625] {};
   \node[ball,text width=.18cm,thick,color=blue,right=.15cm of v7, scale=.625] {};
   \node[ball,text width=.18cm,thick,color=blue,left=.15cm of v6, scale=.625] {};
   \coordinate (a) at ($(v3)-(.125,0)$);
   \coordinate (b) at ($(v3)+(0,.125)$);
   \coordinate (c) at ($(v34)+(0,.175)$);
   \coordinate (d) at ($(v4)+(0,.125)$);
   \coordinate (e) at ($(v4)+(.125,0)$);
   \coordinate (f) at ($(v45)+(.175,0)$);
   \coordinate (g) at ($(v5)+(.125,0)$);
   \coordinate (h) at ($(v5)-(0,.125)$);
   \coordinate (i) at ($(v5)-(.125,0)$);
   \coordinate (j) at ($(v45)-(.175,0)$);
   \coordinate (k) at ($(v34)-(0,.175)$);
   \coordinate (l) at ($(v3)-(0,.125)$);
   \draw [thick, red!50!black] plot [smooth cycle] coordinates {(a) (b) (c) (d) (e) (f) (g) (h) (i) (j) (k) (l)};
   \node[below=.05cm of k, color=red!50!black] {\footnotesize $\displaystyle\mathfrak{g}$};
   \coordinate (n) at ($(v2)+(0,.125)$);
   \coordinate (o) at ($(v2)+(.125,0)$);
   \coordinate (p) at ($(v12)+(.175,0)$);
   \coordinate (q) at ($(v71)+(0,.175)$);
   \coordinate (r) at ($(v7)+(0,.125)$);
   \coordinate (t) at ($(v67)+(0,.175)$);
   \coordinate (ta) at ($(v6)+(0,.125)$);
   \coordinate (tb) at ($(v6)+(.125,0)$);
   \coordinate (tc) at ($(v6)-(0,.125)$);
   \coordinate (td) at ($(v67)-(0,.175)$);
   \coordinate (u) at ($(v71)-(0,.175)$);
   \coordinate (w) at ($(v1)-(0,.125)$);
   \coordinate (x) at ($(v1)-(.125,0)$);
   \coordinate (y) at ($(v12)-(.125,0)$);
   \coordinate (z) at ($(v2)-(.125,0)$);
   \draw [thick, blue] plot [smooth cycle] coordinates {(n) (o) (p) (q) (r) (t) (ta) (tb) (tc) (td) (u) (w) (x) (y) (z)};
   \node[right=.05cm of p, color=blue] {\footnotesize $\displaystyle\bar{\mathfrak{g}}$};
  \end{scope}
 \end{tikzpicture}
\end{wrapfigure}
For the sake of concreteness, let us refer to the graph drawn here on the left. The vertices
$
 \begin{tikzpicture}[ball/.style = {circle, draw, align=center, anchor=north, inner sep=0}]
   \node[ball,text width=.18cm,thick,color=red!50!black] at (0,0) {};
 \end{tikzpicture}
$
correspond to a further scattering facet reached as $x_3+x_4+x_5+y_{23}+y_{37}+y_{56}\longrightarrow0$, $y_{ij}$ being the energies associated to the edges between the vertices $i$ and $j$ -- at this boundary, the energy conservation for $\mathfrak{g}$ is restored.
Notice further that the vertices
$
 \begin{tikzpicture}[ball/.style = {circle, draw, align=center, anchor=north, inner sep=0}]
  \node[ball,text width=.18cm,thick,color=blue] at (0,0) {};
 \end{tikzpicture}
$
are also related to a smaller scattering facet for the subgraph $\bar{\mathfrak{g}}$. Interestingly, the remaining vertices
$
 \begin{tikzpicture}[ball/.style = {circle, draw, align=center, anchor=north, inner sep=0}]
   \node[ball,text width=.18cm,thick,color=red] at (0,0) {};
 \end{tikzpicture}
$
are on the edges connecting $\mathfrak{g}$ and $\bar{\mathfrak{g}}$, {\it i.e.} they are related to the {\it cut edges} connected two lower-point scattering amplitudes. They are the only vertices depending on the vectors of such edges and form a simplex in a lower-dimensional space.

Thus, as $x_3+x_4+x_5+y_{23}+y_{37}+y_{56}\longrightarrow0$, the related face of the scattering facet has the structure of a product of two lower-dimensional scattering facets times the simplex for the vertices related to the edges connecting the two subgraphs. This is a general feature of any face of the scattering facet. Consequently, the canonical form related to such faces is given by
\begin{equation}\label{eq:CFunit}
 \Omega\:=\:
  \left(
   \prod_{e\in\slashed{\mathcal{E}}}\frac{1}{2y_e}
  \right)
  \mathcal{A}[\mathfrak{g}]\times\mathcal{A}[\bar{\mathfrak{g}}],
\end{equation}
where $\slashed{\mathcal{E}}$ is the set of cut edges.

This is a completely general phenomenon: the faces of the scattering facet factorize into a simplex associated with the cut edges, and a product of lower scattering facets. This product can have more than two pieces. For instance let us consider a different subgraph $\mathfrak{g}$, which corresponds to $x_3+x_7+y_{23}+y_{34}+y_{71}+y_{67}\longrightarrow0$.
\begin{wrapfigure}{l}{3.75cm}
 \begin{tikzpicture}[ball/.style = {circle, draw, align=center, anchor=north, inner sep=0}, cross/.style={cross out, draw, minimum size=2*(#1-\pgflinewidth), inner sep=0pt, outer sep=0pt}, scale=1.25, transform shape]
  \begin{scope}
   \coordinate[label=below:{\tiny $x_1$}] (v1) at (0,0);
   \coordinate[label=above:{\tiny $x_2$}] (v2) at ($(v1)+(0,1.25)$);
   \coordinate[label=above:{\tiny $x_3$}] (v3) at ($(v2)+(1,0)$);
   \coordinate[label=above:{\tiny $x_4$}] (v4) at ($(v3)+(1,0)$);
   \coordinate[label=right:{\tiny $x_5$}] (v5) at ($(v4)-(0,.625)$);
   \coordinate[label=below:{\tiny $x_6$}] (v6) at ($(v5)-(0,.625)$);
   \coordinate[label=below:{\tiny $x_7$}] (v7) at ($(v6)-(1,0)$);
   \draw[thick] (v1) -- (v2) -- (v3) -- (v4) -- (v5) -- (v6) -- (v7) -- cycle;
   \draw[thick] (v3) -- (v7);
   \draw[fill=black] (v1) circle (2pt);
   \draw[fill=black] (v2) circle (2pt);
   \draw[fill=black] (v3) circle (2pt);
   \draw[fill=black] (v4) circle (2pt);
   \draw[fill=black] (v5) circle (2pt);
   \draw[fill=black] (v6) circle (2pt);
   \draw[fill=black] (v7) circle (2pt);
   \coordinate (v12) at ($(v1)!0.5!(v2)$);
   \coordinate (v23) at ($(v2)!0.5!(v3)$);
   \coordinate (v34) at ($(v3)!0.5!(v4)$);
   \coordinate (v45) at ($(v4)!0.5!(v5)$);
   \coordinate (v56) at ($(v5)!0.5!(v6)$);
   \coordinate (v67) at ($(v6)!0.5!(v7)$);
   \coordinate (v71) at ($(v7)!0.5!(v1)$);
   \coordinate (v37) at ($(v3)!0.5!(v7)$);
   \node[ball,text width=.18cm,thick,color=red!50!black,below=.15cm of v3, scale=.625] {};
   \node[ball,text width=.18cm,thick,color=red,left=.15cm of v4, scale=.625] {};
   \node[ball,text width=.18cm,thick,color=green!50!black,below=.1cm of v4, scale=.625] {};
   \node[ball,text width=.18cm,thick,color=green!50!black,above=.1cm of v5, scale=.625] {};
   \node[ball,text width=.18cm,thick,color=red,right=.15cm of v2, scale=.625] {};
   \node[ball,text width=.18cm,thick,color=red!50!black,above=.15cm of v7, scale=.625] {};
   \node[ball,text width=.18cm,thick,color=green!50!black,above=.1cm of v6, scale=.625] {};
   \node[ball,text width=.18cm,thick,color=blue,below=.15cm of v2, scale=.625] {};
   \node[ball,text width=.18cm,thick,color=blue,above=.15cm of v1, scale=.625] {};
   \node[ball,text width=.18cm,thick,color=red,right=.15cm of v1, scale=.625] {};
   \node[ball,text width=.18cm,thick,color=green!50!black,below=.15cm of v5, scale=.625] {};
   \node[ball,text width=.18cm,thick,color=red,left=.15cm of v6, scale=.625] {};
   \coordinate (a) at ($(v3)-(.125,0)$);
   \coordinate (b) at ($(v3)+(0,.125)$);
   \coordinate (c) at ($(v3)+(.125,0)$);
   \coordinate (d) at ($(v37)+(.175,0)$);
   \coordinate (e) at ($(v7)+(.125,0)$);
   \coordinate (f) at ($(v7)-(0,.125)$);
   \coordinate (g) at ($(v7)-(.125,0)$);
   \coordinate (h) at ($(v37)-(.175,0)$);
   \draw [thick, red!50!black] plot [smooth cycle] coordinates {(a) (b) (c) (d) (e) (f) (g) (h)};
   \node[right=.05cm of d, color=red!50!black] {\footnotesize $\displaystyle\mathfrak{g}$};
   \coordinate (al) at ($(v2)-(.125,0)$);
   \coordinate (bl) at ($(v2)+(0,.125)$);
   \coordinate (cl) at ($(v2)+(.125,0)$);
   \coordinate (dl) at ($(v12)+(.175,0)$);
   \coordinate (el) at ($(v1)+(.125,0)$);
   \coordinate (fl) at ($(v1)-(0,.125)$);
   \coordinate (gl) at ($(v1)-(.125,0)$);
   \coordinate (hl) at ($(v12)-(.175,0)$);
   \draw [thick, blue] plot [smooth cycle] coordinates {(al) (bl) (cl) (dl) (el) (fl) (gl) (hl)};
   \node[left=.05cm of hl, color=blue] {\footnotesize $\displaystyle\mathfrak{g}_{\mbox{\tiny L}}$};
   \coordinate (ar) at ($(v4)-(.125,0)$);
   \coordinate (br) at ($(v4)+(0,.125)$);
   \coordinate (cr) at ($(v4)+(.125,0)$);
   \coordinate (dr) at ($(v5)+(.175,0)$);
   \coordinate (er) at ($(v6)+(.125,0)$);
   \coordinate (fr) at ($(v6)-(0,.125)$);
   \coordinate (gr) at ($(v6)-(.125,0)$);
   \coordinate (hr) at ($(v5)-(.175,0)$);
   \draw [thick, green!50!black] plot [smooth cycle] coordinates {(ar) (br) (cr) (dr) (er) (fr) (gr) (hr)};
   \node[right=.08cm of dr, color=green!50!black] {\footnotesize $\displaystyle\mathfrak{g}_{\mbox{\tiny R}}$};
  \end{scope}
 \end{tikzpicture}
\end{wrapfigure}
As before, the vertices
$
 \begin{tikzpicture}[ball/.style = {circle, draw, align=center, anchor=north, inner sep=0}]
   \node[ball,text width=.18cm,thick,color=red!50!black] at (0,0) {};
 \end{tikzpicture}
$
identify the lower-dimensional scattering facet related to the subgraph $\mathfrak{g}$. The other open circles correspond to the other vertices of the face we are discussing. In particular the vertices
$
 \begin{tikzpicture}[ball/.style = {circle, draw, align=center, anchor=north, inner sep=0}]
   \node[ball,text width=.18cm,thick,color=blue] at (0,0) {};
 \end{tikzpicture}
$
and
$
 \begin{tikzpicture}[ball/.style = {circle, draw, align=center, anchor=north, inner sep=0}]
   \node[ball,text width=.18cm,thick,color=green!50!black] at (0,0) {};
 \end{tikzpicture}
$
separately identify a further lower-dimensional scattering facet each. Finally, the vertices
$
 \begin{tikzpicture}[ball/.style = {circle, draw, align=center, anchor=north, inner sep=0}]
   \node[ball,text width=.18cm,thick,color=red] at (0,0) {};
 \end{tikzpicture}
$
mark the cut edges. Thus, the canonical form related to this face of the scattering facet is
\begin{equation}\label{eq:CFunit2}
 \Omega\:=\:
  \left(
   \prod_{e\in\slashed{\mathcal{E}}}\frac{1}{2y_e}
  \right)
  \mathcal{A}[\mathfrak{g}]\times\mathcal{A}[\mathfrak{g}_{\mbox{\tiny L}}]\times\mathcal{A}[\mathfrak{g}_{\mbox{\tiny R}}].
\end{equation}

Now, this geometric factorization of the boundaries of the facet implies that the residues of the poles for the (integrand of) scattering amplitudes factorize in exactly the same way. This is a basic and fundamental statement about the integrand, which implies the familiar cutting rules that ensure unitarity in perturbation theory. The usual statement of unitarity for the $S$-matrix $S=1 - i T$, is
\begin{equation*}
 \begin{split}
  -i \langle out |(T - T^\dagger)| in \rangle &\:=\:\langle out |T^\dagger T | in \rangle\:=\\
                                              &\:=\:\sum_I \langle out|T^\dagger|I\rangle \langle I| T |in \rangle.
 \end{split}
\end{equation*}
The imaginary parts in the left-hand-side arise from the Feynman $i \epsilon$ in the usual way as Im$\left\{1/(E - i \epsilon)\right\} = \delta(E)$. Thus a given residue of the scattering amplitude integrand computes a contribution to the imaginary part of the amplitude. Now the factors of factors $1/2y_e$ coming from the simplex associated with the cut edges,  together with $d^d \mathbf{\ell}$, form the Lorentz invariant phase-space for the intermediate lines, while the factors associated with the smaller scattering facets gives us the correct factorization into the product of lower amplitudes. Note that the cases involving more than one lower scattering facet factor correctly include disconnected components in the computation of $\langle out | T^\dagger T | in \rangle$.  Integrating the product of the two tree-level amplitudes over such phase-space computes the discontinuity along a branch cut which is the imaginary part of the loop amplitude. Note also that, importantly,
the open circles do not just identify the vertices of the face and its structure, but provide the crucial information about the direction of the flow of the energy. Consider any of the subgraphs we identify for a given face. Then, if an open circle
$
 \begin{tikzpicture}[ball/.style = {circle, draw, align=center, anchor=north, inner sep=0}]
   \node[ball,text width=.18cm,thick,color=red] at (0,0) {};
 \end{tikzpicture}
$
is far away from the subgraph, the energy associated to the edge marked by it, is incoming. If instead such an open circle is closer, the energy of the related edge is outgoing -- this can be easily seen from the fact that such energies $y_e$ appear with positive/negative sign respectively.


\section{Emergent Lorentz invariance}\label{sec:ELI}

Let us now turn to seeing how Lorentz invariance emerges from the scattering facet, where the challenge is clear. Ordinarily, the Lorentz invariance of scattering amplitudes is manifested by expressing the amplitudes as a function of Lorentz-invariant kinematical variables like Mandelstam invariants. But the wavefunction only depends on spatial momenta and associated energy variables. Already at tree level, we must understand how, sitting on the total energy pole, the poles associated with energy denominators, as seen in the wavefunction, pair-up into Lorentz-invariant propagators. The story at loop level must be even more interesting, since at the level of the integrand, while we see the spatial loop momenta, we don't have an analog of ``$l_0$'', the time component; to make Lorentz invariance manifest we must somehow see these ``extra" variables corresponding to $l_0$ appear, and again have the fundamentally-linear-in-energy poles pair up into Lorentz-invariant propagators.

The emergence of Lorentz-invariance at tree-level was already considered in \cite{Arkani-Hamed:2017fdk}, so let us begin by reviewing this story. Tree-level graphs have $n_v\,=\,n_e+1$ vertices. The related scattering facet has, thus, $2n_e\,\equiv\,n_e+n_v-1$ and, consequently, it is a $(n_e+n_v-2)$-dimensional polytope with $n_e+n_v-1$ vertices, {\it i.e.}  it is a simplex $\mathcal{S}$.

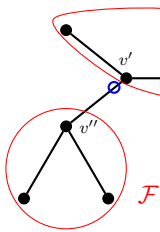
\begin{wrapfigure}{l}{4cm}
 \begin{tikzpicture}[ball/.style = {circle, draw, align=center, anchor=north, inner sep=0}, cross/.style={cross out, draw, minimum size=2*(#1-\pgflinewidth), inner sep=0pt, outer sep=0pt}, scale=.8, transform shape]
  \node[ball,text width=.18cm,fill,color=black,label=above:{\footnotesize $v'$}] at (0,0) (x1) {};
  \node[ball,text width=.18cm,fill,color=black,right=1.2cm of x1.east] (x2) {};
  \node[ball,text width=.18cm,fill,color=black,right=1.2cm of x2.east] (x3) {};
  \node[ball,text width=.18cm,fill,color=black] at (-1,.8) (x4) {};
  \node[ball,text width=.18cm,fill,color=black,label=right:{\footnotesize $v''$}] at (-1,-.8) (x5) {};
  \node[ball,text width=.18cm,fill,color=black] at (-1.7,-2) (x6) {};
  \node[ball,text width=.18cm,fill,color=black] at (-.3,-2) (x7) {};

  \node[above=.35cm of x5.north] (ref2) {};
  \coordinate (Int2) at (intersection of x5--x1 and ref2--x2);

  \coordinate (t1) at (x3.east);
  \coordinate (t2) at (x4.west);
  \coordinate (t3) at (x1.south west);
  \coordinate (t4) at (x2.south);

  \draw[-,thick,color=black] (x1) -- (x2) -- (x3);
  \draw[-,thick,color=black] (x1) -- (x4);
  \draw[-,thick,color=black] (x5) -- (x1);
  \draw[-,thick,color=black] (x5) -- (x7);
  \draw[-,thick,color=black] (x5) -- (x6);
  \node[ball,text width=.18cm,thick,color=blue,left=-.09cm of Int2] {};
  \draw[red] plot [smooth cycle] coordinates {(3,-.1) (1.2,1) (-1.2,.9) (t3) (1.5,-.5)};
  \node[color=red,right=.3cm of x3.east] {\large $\mathcal{C}$};

  \draw[red] (-1,-1.6) circle (1cm);
  \node[color=red,right=.3cm of x7.north east] {\large $\mathcal{F}$};
 \end{tikzpicture}
\end{wrapfigure}

For each edge $e$ of $\mathcal{G}$, it has two vertices and a facet of $\mathcal{S}$ has just one of them: The facet can then be identified by further marking the graph with an open circle for the vertex which belongs to it. This marking splits the graphs in two parts, $\mathcal{C}$ and $\mathcal{F}$. The hyperplane associated to any of the vertices of this facet of $\mathcal{S}$ can be written as $\omega={\bf \tilde{Y}}_e+\sum_{v\in\mathcal{C}}{\bf\tilde{X}}_v$, and annihilates all the vertices associated to $\mathcal{F}$, the vertices related to edges that do not touch any of the vertices of $e$, the ones associated to the edges of $\mathcal{C}$ touching $v'$, as well as the vertex related to the uncircled end of $e$. Then, $\omega\cdot\mathcal{Y}=y_e+\sum_{v\in\mathcal{C}}x_v$. Finally, we can circle the other end of the same edge, for which the same discussion holds and $\omega'\cdot\mathcal{Y}=y_e+\sum_{v\in\mathcal{F}}x_v$. However, on $\mathcal{S}$, $\sum_vx_v\,=\,0$ and therefore $\omega'\cdot\mathcal{Y}=y_e-\sum_{v\in\mathcal{C}}x_v$. Thus the canonical form for $\mathcal{S}$ is
\begin{equation}\label{eq:LorTree}
 \Omega(\mathcal{S})\:=\:\prod_{e\in\mathcal{E}_{\mathcal{S}}}\frac{1}{y_e^2-\left(\sum_{v\in\mathcal{C}_e}x_v\right)^2}
\end{equation}
which is nothing but the product of Lorentz invariant propagators!

Let us now move on to discussing Lorentz-invariance at loop level. As alluded to above, one basic challenge is to see where the ``$l_0$" variables (needed for manifestly Lorentz-invariant loop integrands) will come from. This turns out to have a beautiful answer, related to a representation for the canonical form for general polytopes that has played a prominent role in a number of other settings. Given any projective polyotpe ${\cal P}$ in a projective space ${\cal Y}$,  its associated canonical form $\Omega({\cal Y};{\cal P})$ can be determined by a contour integral \cite{Arkani-Hamed:2017tmz}:
\begin{equation}\label{eq:CFci}
 \Omega(\mathcal{Y};\mathcal{P})\:=\:\int_{\mathbb{R}^N}\prod_{j=1}^{\nu}\frac{dc_j}{c_j-i\varepsilon_j}\delta^{\mbox{\tiny $(N)$}}\left(\mathcal{Y}-\sum_{j=1}^{\nu}c_j{\bf V}^{\mbox{\tiny $(j)$}}\right)
\end{equation}
where $\nu$ is the number of vertices of $\mathcal{P}$.  This formula can be obtained via Fourier/Laplace transforms, of another important representation of $\Omega({\cal Y};{\cal P})$, which identifies $\Omega$ with the volume of the {\it dual} polytope $\tilde{P}$, relative to ${\cal Y}$ as the hyperplane at infinity.

We now specialize this representation of the canonical form for the scattering facet of a cosmological polytope $\mathcal{P}_{\mathcal{G}}$ related to an arbitrary $L$-loop graph $\mathcal{G}$. In this case, $\mathcal{G}$ has $n_e$ edges and $n_v\,=\,n_e+1-L$ vertices. Then the scattering facet lives in $\mathbb{P}^{n_e+n_v-2}$ and has $2n_e=n_e+n_v-1+L$ vertices: it has $L$ vertices more than a simplex, which it reduces to just for $L=0$ ({\it i.e.} at tree level). Thus, the canonical form for  $\mathcal{P}_{\mathcal{G}}$ is represented by \eqref{eq:CFci} provided that $N\equiv n_e+n_v-1$ and $\nu\equiv n_e+n_v-1+L$. Starting with such a representation for the scattering facet, one can observe that the $\delta$ functions localize $n_e+n_v-1$ integration variables, leaving unfixed exactly $L$ of them. Indeed, there is some freedom in choosing the unfixed $c$'s: each of such choices corresponds to select all those hyperplanes in the scattering facet that do not contain the vertices associated to the unfixed $c$'s.
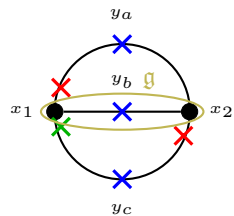
\begin{wrapfigure}{l}{3cm}
 \begin{tikzpicture}[ball/.style = {circle, draw, align=center, anchor=north, inner sep=0}, cross/.style={cross out, draw, minimum size=2*(#1-\pgflinewidth), inner sep=0pt, outer sep=0pt}, scale=1.2, transform shape]
  \coordinate (cnt) at (0,0);
  \draw[thick, color=black] (cnt) circle (.75cm);
  \node[ball, text width=.18cm,fill,color=black, label=left:$\mbox{\tiny $x_1$}$, left=.65cm of cnt] (x1) {};
  \node[ball, text width=.18cm,fill,color=black, label=right:$\mbox{\tiny $x_2$}$, right=.65cm of cnt] (x2) {};
  \draw[-, thick] (x1) -- node[label=above:{\tiny $y_b$}] {} (x2);
  \node[label=above:$\mbox{\tiny $y_{a}$}$, above=.65cm of cnt] {};
  \node[label=below:$\mbox{\tiny $y_{c}$}$, below=.65cm of cnt] {};
  \node[very thick, cross=4pt, rotate=0, color=blue, above=.65cm of cnt]{};
  \node[very thick, cross=4pt, rotate=0, color=blue] at (cnt){};
  \node[very thick, cross=4pt, rotate=0, color=blue, below=.65cm of cnt]{};

  \node[very thick, cross=4pt, rotate=0, color=red, above=.1cm of x1.north east] {};
  \node[very thick, cross=4pt, rotate=0, color=red, below=.1cm of x2.south west] {};
  \node[very thick, cross=4pt, rotate=0, color=green!70!black, below=.cm of x1.south east]{};

  \draw[thick, color=yellow!70!black] (cnt) ellipse (.9cm and .2cm);
  \node[color=yellow!70!black] (C) at ($(cnt)+(.3,.35)$) {\footnotesize $\mathfrak{g}$};
 \end{tikzpicture}
\end{wrapfigure}
We can indicate such vertices with red crosses on the graph $\mathcal{G}$. Each codimension-one hyperplane which does not contain the marked vertices, is then identified by further marking the graph with a green cross for the vertex which does not belong to it, as the subgraph $\mathfrak{g}$ enclosed within the red/green crosses $\omega=\sum_{v\in\mathfrak{g}}{\bf\tilde{X}}_v+\sum_{e\in\mathcal{E}_{\mathfrak{g}}^c}{\bf\tilde{Y}}_e-\sum_{e\in\mathcal{E}_{\mathfrak{g}}^u}{\bf\tilde{Y}}_e$, with $\mathcal{E}_{\mathfrak{g}}^c$ and $\mathcal{E}_{\mathfrak{g}}^u$ being the sets of external edges of $\mathfrak{g}$ with and without cross respectively. Defining $\mathcal{V}_r$ as the set of vertices marked with the red crosses ({\it i.e.} those ones whose related $c$'s are unfixed) and $\tilde{\mathcal{Y}}\equiv\mathcal{Y}-\left(\sum_{\mbox{\tiny $j\in\mathcal{V}_r$}}c_j{\bf V}^{\mbox{\tiny $(j)$}}\right)$, then $\omega\cdot\tilde{\mathcal{Y}}$ is the solution for $c$ related to the vertex marked with the green cross. Notice that if both ends of an edge $\bar{e}$ are red/green crossed, then $\omega$ annihilates all the ${\bf V}^{\mbox{\tiny $(j)$}}$ in $\mathcal{V}_r$ but one, and $\omega\cdot\mathcal{Y}\sim2y_{\bar{e}}$. Hence, these contributions provide a factor of the form $\prod_{j=1}^L(c_j-y_{e_j}+i\varepsilon_{k_j})^{-1}$, $c_j$'s being the unfixed variables. Furthermore, for all the other edges, the green cross can mark either of the two ends and, thus, the two contributions depend on the same $x$'s and $y$'s and they differ only for the sign of the $y$ related to the edge in question. Thus, the canonical form for the scattering facet acquires the form
\begin{widetext}
\begin{equation}\label{eq:CFlp}
  \Omega(\mathcal{P}_{\mathcal{G}})\:=\:\int\prod_{j=1}^L\frac{dc_j}{\left(c_j-\frac{y_{e_j}}{2}\right)^2-\left(\frac{y_{e_j}}{2}-i\varepsilon_j\right)^2}
  \prod_{s=1}^{n_e-L}\frac{1}{\left(\sum_{r}\sigma_{r_s}c_{r}-\frac{\mathfrak{y}_s}{2}\right)^2-\left(\frac{y_s}{2}-i\varepsilon_s\right)^2}
\end{equation}
\end{widetext}
where $\sigma_{r_s}$ are suitable signs, $\mathfrak{y}_s$ are combinations of $y$'s and $x$'s. Each quadratic factor above is a Lorentz invariant propagator, with the $c_j$'s which are nothing but the $l_0$ component of the loop momenta! Adding the $d$-dimensional loop measures, $d^d\mathbf{\ell}^{\mbox{\tiny $(j)$}}$, the canonical form \eqref{eq:CFlp} returns a Lorentz invariant loop integrand for the related graph. It is remarkable how the contour integral representation makes Lorentz invariance manifest, with the propagators inheriting the correct $i\varepsilon$ prescription from the standard one for the canonical form!

Finally, we can also perform the contour integration over the leftover $c_j$'s, for which we have the freedom to close integration contours in a number of ways. From the canonical form perspective, these different ways of closing the integration contours return all the possible triangulations for the polytope \cite{Arkani-Hamed:2017tmz}. In our case, we have just learnt that these integrations are exactly the $l_0$ integration and, thus, all the possible ways of triangulating the scattering facet correspond to all the possible representations obtainable by performing the $l_0$ integrations by contour integration using the $i \epsilon$ poles (as familiar from the ``Feynman tree theorem" \cite{Feynman:1963ax, CaronHuot:2010zt}), in all possible ways! As a visualizable example, let's consider the scattering facet of the two-loop two-site graph, which is a truncated tetrahedron in $\mathbb{P}^3$:
\begin{widetext}
\begin{equation*}
\centering
\begin{tikzpicture}[ball/.style = {circle, draw, align=center, anchor=north, inner sep=0}, scale={.75}, transform shape]
  \begin{scope}[scale={1.25}, transform shape]
   \coordinate (cnt) at (0,0);
   \draw[thick, color=black] (cnt) circle (.75cm);
   \node[ball, text width=.18cm,fill,color=black, label=left:$\mbox{\large $x_1$}$, left=.65cm of cnt] (x1) {};
   \node[ball, text width=.18cm,fill,color=black, label=right:$\mbox{\large $x_2$}$, right=.65cm of cnt] (x2) {};
   \draw[-, thick] (x1) -- node[label=above:{\large $y_c$}] {} (x2);
   \node[label=above:$\mbox{\large $y_{a}$}$, above=.65cm of cnt] {};
   \node[label=below:$\mbox{\large $y_{b}$}$, below=.65cm of cnt] {};
   \node[right=1.5cm of x2, scale=.8] (v2) {\large $\displaystyle {\bf x}_1+{\bf y}_{c}-{\bf x}_2,\;-{\bf x}_1+{\bf y}_{c}+{\bf x}_2$};
   \node[above=.3cm of v2, scale=.8] (v1) {\large $\displaystyle \left\{{\bf x}_1+{\bf y}_{a}-{\bf x}_2,\;-{\bf x}_1+{\bf y}_{a}+{\bf x}_2\right.$};
   \node[below=.3cm of v2, scale=.8] (v3) {\large $\displaystyle \left.{\bf x}_1+{\bf y}_{b}-{\bf x}_2,\;-{\bf x}_1+{\bf y}_{b}+{\bf x}_2\right\}$};
  \end{scope}

  \begin{scope}[shift={(10,-1.75)}, transform shape]
   \def\r{2.5}
   \pgfmathsetmacro\ax{\r*cos(30)}
   \pgfmathsetmacro\ay{\r*sin(30)}
   \coordinate[label=below:{\large ${\bf 3}$}] (v3) at (0,0);
   \coordinate[label=above:{\large ${\bf 1}$}] (v1) at ($(v3)+(0,3)$);
   \coordinate[label=below:{\large ${\bf 5}$}] (v5) at ($(v3)+(3,0)$);
   \coordinate[label=left:{\large ${\bf 4}$}] (v4) at ($(v3)+(\ax,\ay)$);
   \coordinate[label=above:{\large ${\bf 2}$}] (v2) at ($(v1)+(\ax,\ay)$);
   \coordinate[label=right:{\large ${\bf 6}$}] (v6) at ($(v5)+(\ax,\ay)$);

   \draw[-,dashed,fill=red!40, opacity=.6] (v3) -- (v4) -- (v6) -- (v5) -- cycle;
   \draw[-,dashed,fill=blue!40, opacity=.6] (v3) -- (v4) -- (v2) -- (v1) -- cycle;
   \draw[-,dashed,fill=red!60, opacity=.2] (v6) -- (v4) -- (v2) -- cycle;
   \draw[-,thick, fill=blue!60, opacity=.2] (v3) -- (v5) -- (v1) -- cycle;
   \draw[-,thick, fill=yellow!60!black, opacity=.5] (v1) -- (v2) -- (v6) -- (v5) -- cycle;
  \end{scope}
 \end{tikzpicture}
\end{equation*}
\end{widetext}
\noindent
where the labels $\{{\bf 1}, {\bf 2}, {\bf 3}, {\bf 4}, {\bf 5}, {\bf 6}\}$ identify the vertices as they appear in the list above.
It has six possible triangulations, which is exactly the number of ways the two $l_0$ integrations can be performed: once the order of integration is chosen, each of the two integrations can be performed in the upper or lower half plane, returning four representations, while changing the order of integration, one finds two more inequivalent representations. The canonical form of the scattering facet, as represented by the following triangulation
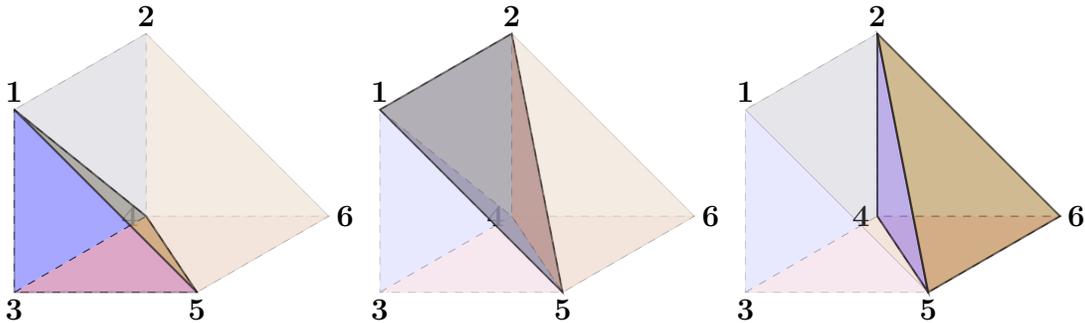
\begin{widetext}
\begin{equation*}\label{eq:Tr45}
 \begin{tikzpicture}[ball/.style = {circle, draw, align=center, anchor=north, inner sep=0}, scale={.9}, transform shape]
  \begin{scope}[scale={.9}, transform shape]
   \def\r{2.5}
   \pgfmathsetmacro\ax{\r*cos(30)}
   \pgfmathsetmacro\ay{\r*sin(30)}
   \coordinate[label=below:{\Large ${\bf 3}$}] (v3) at (0,0);
   \coordinate[label=above:{\Large ${\bf 1}$}] (v1) at ($(v3)+(0,3)$);
   \coordinate[label=below:{\Large ${\bf 5}$}] (v5) at ($(v3)+(3,0)$);
   \coordinate[label=left:{\Large ${\bf 4}$}] (v4) at ($(v3)+(\ax,\ay)$);
   \coordinate[label=above:{\Large ${\bf 2}$}] (v2) at ($(v1)+(\ax,\ay)$);
   \coordinate[label=right:{\Large ${\bf 6}$}] (v6) at ($(v5)+(\ax,\ay)$);

   \draw[-,dashed,fill=red!40, opacity=.15] (v3) -- (v4) -- (v6) -- (v5) -- cycle;
   \draw[-,dashed,fill=blue!40, opacity=.15] (v3) -- (v4) -- (v2) -- (v1) -- cycle;
   \draw[-,dashed,fill=red!60, opacity=.05] (v6) -- (v4) -- (v2) -- cycle;
   \draw[-,fill=blue!60, opacity=.05] (v3) -- (v5) -- (v1) -- cycle;
   \draw[-,fill=yellow!60!black, opacity=.125] (v1) -- (v2) -- (v6) -- (v5) -- cycle;

   \draw[-,dashed,fill=red!40, opacity=.6] (v3) -- (v4) -- (v5) -- cycle;
   \draw[-,dashed,fill=blue!40, opacity=.6] (v3) -- (v4) -- (v1) -- cycle;
   \draw[-,thick,fill=yellow!60!black, opacity=.5] (v4) -- (v5) -- (v1) -- cycle;
   \draw[-,thick,fill=blue!60, opacity=.2] (v1) -- (v3) -- (v5) -- cycle;
  \end{scope}

  \begin{scope}[scale={.9}, shift={(6,0)}, transform shape]
   \def\r{2.5}
   \pgfmathsetmacro\ax{\r*cos(30)}
   \pgfmathsetmacro\ay{\r*sin(30)}
   \coordinate[label=below:{\Large ${\bf 3}$}] (v3) at (0,0);
   \coordinate[label=above:{\Large ${\bf 1}$}] (v1) at ($(v3)+(0,3)$);
   \coordinate[label=below:{\Large ${\bf 5}$}] (v5) at ($(v3)+(3,0)$);
   \coordinate[label=left:{\Large ${\bf 4}$}] (v4) at ($(v3)+(\ax,\ay)$);
   \coordinate[label=above:{\Large ${\bf 2}$}] (v2) at ($(v1)+(\ax,\ay)$);
   \coordinate[label=right:{\Large ${\bf 6}$}] (v6) at ($(v5)+(\ax,\ay)$);

   \draw[-,dashed,fill=red!40, opacity=.15] (v3) -- (v4) -- (v6) -- (v5) -- cycle;
   \draw[-,dashed,fill=blue!40, opacity=.15] (v3) -- (v4) -- (v2) -- (v1) -- cycle;
   \draw[-,dashed,fill=red!60, opacity=.05] (v6) -- (v4) -- (v2) -- cycle;
   \draw[-,fill=blue!60, opacity=.05] (v3) -- (v5) -- (v1) -- cycle;
   \draw[-,fill=yellow!60!black, opacity=.125] (v1) -- (v2) -- (v6) -- (v5) -- cycle;

   \draw[-,dashed,fill=blue!40, opacity=.6] (v1) -- (v2) -- (v5) -- cycle;
   \draw[-,dashed,fill=red!60, opacity=.3] (v5) -- (v4) -- (v2) -- cycle;
   \draw[-,dashed,fill=blue!60, opacity=.3] (v4) -- (v5) -- (v1) -- cycle;
   \draw[-,thick,fill=yellow!60!black, opacity=.4] (v1) -- (v2) -- (v5) -- cycle;
  \end{scope}

  \begin{scope}[scale={.9}, shift={(12,0)}, transform shape]
   \def\r{2.5}
   \pgfmathsetmacro\ax{\r*cos(30)}
   \pgfmathsetmacro\ay{\r*sin(30)}
   \coordinate[label=below:{\Large ${\bf 3}$}] (v3) at (0,0);
   \coordinate[label=above:{\Large ${\bf 1}$}] (v1) at ($(v3)+(0,3)$);
   \coordinate[label=below:{\Large ${\bf 5}$}] (v5) at ($(v3)+(3,0)$);
   \coordinate[label=left:{\Large ${\bf 4}$}] (v4) at ($(v3)+(\ax,\ay)$);
   \coordinate[label=above:{\Large ${\bf 2}$}] (v2) at ($(v1)+(\ax,\ay)$);
   \coordinate[label=right:{\Large ${\bf 6}$}] (v6) at ($(v5)+(\ax,\ay)$);

   \draw[-,dashed,fill=red!40, opacity=.15] (v3) -- (v4) -- (v6) -- (v5) -- cycle;
   \draw[-,dashed,fill=blue!40, opacity=.15] (v3) -- (v4) -- (v2) -- (v1) -- cycle;
   \draw[-,dashed,fill=red!60, opacity=.05] (v6) -- (v4) -- (v2) -- cycle;
   \draw[-,fill=blue!60, opacity=.05] (v3) -- (v5) -- (v1) -- cycle;
   \draw[-,fill=yellow!60!black, opacity=.125] (v1) -- (v2) -- (v6) -- (v5) -- cycle;

   \draw[-,dashed,fill=red!40, opacity=.6] (v4) -- (v5) -- (v6) -- cycle;
   \draw[-,dashed,fill=red!60, opacity=.2] (v2) -- (v4) -- (v6) -- cycle;
   \draw[-,thick,fill=blue!40, opacity=.6] (v4) -- (v5) -- (v2) -- cycle;
   \draw[-,thick,fill=yellow!60!black, opacity=.5] (v5) -- (v6) -- (v2) -- cycle;
  \end{scope}
 \end{tikzpicture}
\end{equation*}
\end{widetext}
corresponds to perform both $l_0$ integrations in the upper half plane:

\begin{equation}\label{eq:CF45}
   \begin{split}
    \Omega\:=\:&\frac{1}{2y_a}\frac{1}{2y_{b}}\frac{1}{y_c^2-(y_a+y_b-x_1)^2}\:+\\
            &+\frac{1}{2y_b}\frac{1}{2y_c}\frac{1}{y_a^2-(y_b-y_c-x_1)^2}\:+\\
            &+\:\frac{1}{2y_c}\frac{1}{2y_a}\frac{1}{y_b^2-(y_c+y_a+x_1)^2},
   \end{split}
\end{equation}
where the terms in the sum corresponds to the simplices in the order as shown in the picture above. This expression precisely corresponds to one way of performing the $l_0$ integrals using contour integration; with the $1/(2y)$ factors arising from the residues on the $i\epsilon$ poles. 


\section{Outlook}\label{sec:CO}

Cosmological polytopes have an extremely simple intrinsic definition, fully described by a collection of triangles, each of which is allowed to intersect other triangles on only two of its three sides. Any such collection of intersecting triangles is naturally associated with a graph, and the convex hull of the triangle vertices given the cosmological polytope. The canonical form of the polytope computes the contribution to the (integrand of) wavefunction from the corresponding graph. The singularities of the wavefunction are reflected  in the facets of the polytope, and one of these is the {\it scattering facet} associated with the total energy pole, which gives flat-space scattering amplitudes.  We have seen how the exact Lorentz invariance and unitarity of the amplitudes arises in a simple way from the  geometry of the scattering facet. It is rather remarkable to see the fundamental rules of both cosmology and particle scattering arising in a simple way from such primitive, essentially combinatorial ideas. These results suggest a number of obvious directions for future exploration.

The cosmological polytope gives a direct understanding of the wavefunction, but it is natural to ask whether there is a similar geometric object that directly computes the squared modulus of the wavefunction, or directly, the spatial correlation functions \cite{Arkani-Hamed:2018ahp}. Also, the scattering facet is of fundamental importance to the full geometry of the polytope; indeed, general facets of the cosmological polytope factorize into products of lower cosmological polytopes and scattering facets. This suggests that we should be able to find an understanding of both the geometry and the wavefunction itself, building on the scattering facet/amplitude. This theme will be taken up in \cite{Benincasa:2018bp3}.

As also stressed in \cite{Arkani-Hamed:2017fdk}, cosmological polytopes only connect combinatorial geometry to physics one graph at a time. This  is very unsatisfying, especially when contrasted with the story of amplituhedra/associahedra, where  all the magic is in how these structures replace the sum over {\it all} diagrams with new ideas. It is natural to try and connect these deeper ideas to cosmology; the closest point of connection to our current explorations should be to the story of the associahedron for biadjoint $\phi^3$ theory.  There, each Feynman graph for the amplitude corresponds to a simplex in a natural triangulation of the associahedron, so we might try to find a larger cousin of the associahedron, for which the cosmological polytopes play the role of ``simplices".  Indeed our study of the scattering facets in this letter suggest looking for an even more immediate connection. We have seen that the scattering facet of cosmological polytopes gives a new geometric picture {\it for the amplitude itself}, where the Lorentz-invariant poles of amplitudes -- which are quadratic in momenta -- are resolved more fundamentally into products of poles that are linear in energy variables. This suggests that it should be possible to ``double" the associahedron itself, lifting it to a space with roughly twice as many facets, where all propagators are resolved into a pair of poles in this way. This uplift of the associahedron should correspond to the {\it scattering facet} of a conjectural {\it cosmological associahedron}.

It would be very interesting to see whether such a cosmological associahedron exists. Apart from its intrinsic interest, we have understood that the usual associahedron is very closely connected to the physics of the string worldsheet, and indeed the ``kinematic" associahedron provides a natural bridge between worldsheet and space-time physics. It would be fascinating to find and explore the analog of this connection in cosmology.
\\

\acknowledgments

\paragraph{Acknowledgments}-- P.B. is grateful to the Institute for Advanced Study for hospitality during several stages of this work. Some of the polytope analysis has been performed with the aid of {\tt polymake} \cite{polymake:2000}, and the triangulation studies with {\tt TOPCOM} \cite{Rambau:TOPCOM-ICMS:2002}. The work of N.A-H. is supported by DOE under grant DOE DE-SC0009988. P.B. is supported in part by a grant from the Villum Fonden, an ERC-StG grant (N. 757978) and the Danish National Research Foundation (DNRF91).

\bibliographystyle{JHEP}
\bibliography{cprefs}

\end{document}